\documentclass[12pt,a4paper]{article}
\usepackage[multiple]{footmisc}
\usepackage{amsmath}\usepackage{amsfonts}\usepackage{amssymb}\usepackage{latexsym}
\usepackage{graphicx}
\usepackage{rotating}
\usepackage{csquotes}
\normalsize%%%%%%%%%%%%%%%%%%%%%%%%%%%%%

\usepackage[usenames]{color}
\usepackage[abs]{overpic}

                     \def\ignore#1{} %
\newcommand\euro{{\sffamily C%%    
\makebox[0pt][l]{\kern-.70em\mbox{--}}%%  
\makebox[0pt][l]{\kern-.68em\raisebox{.25ex}{--}}}}\def\euro{\hbox{\def\eurorule##1##2##3{\raise##1ex\hbox{\kern##2em  \vrule height.03ex depth0pt width##3em}}\lcmsseuro\kern.05em  \ooalign{C\cr  \eurorule{.41}0{.455}\cr\eurorule{.44}{.005}{.455}\cr  \eurorule{.47}{.01}{.455}\cr\eurorule{.5}{.015}{.455}\cr  \eurorule{.71}0{.5}\cr\eurorule{.74}{.005}{.5}\cr  \eurorule{.77}{.01}{.5}\cr\eurorule{.8}{.015}{.5}}}}

\font\lcmsseuro=lcmss8 at 9pt 

\font\lcmsseuro=lcmss8 at 10pt

\font\lcmsseuro=lcmss8 at 12pt 

\def\euro{{\sf\texteuro}}                             

\begin{document}
\title{Measures of hydroxymethylation}
\author{Alla Slynko\thanks{Ulm University of Applied Sciences, email: {\tt slynko@hs-ulm.de}} \and\setcounter{footnote}{6} Axel Benner\thanks{German Cancer Research Institute DKFZ, email: {\tt benner@dkfz.de}}}
\maketitle

%%%%%%%%%%%%%%%%%%%%%%%%%%%%%
%%%%%%%%%%%%%%%%%%%%%%%%%%%%%
\begin{abstract}
\noindent Hydroxymethylcytosine (5hmC) methylation  is well-known epigenetic mark impacting genome stability.  In this paper, we address the existing 5hmC measure $\Delta \beta$ and discuss its properties both analytically and empirically on real data. Then we introduce several alternative hydroxymethylation measures and compare their properties with those of $\Delta \beta$.   All results are illustrated by means of real data analyses. 
\end{abstract}

%%%%%%%%%%%%%%%%%%%%%%%%%%%%%
%%%%%%%%%%%%%%%%%%%%%%%%%%%%%
\section{Introduction}
DNA methylation is known to play a crucial role in the development of such diseases as diabetes, schizophrenia, and some forms of cancer; see \cite{li} and references therein. In order to address the possible  impact of DNA methylation on the various biological functions and processes, a whole string of extensive biological, bioinformatical, and statistical analyses was developed in the past years \cite{li}. A substantial part of the methods introduced in those analyses aims at quantifying the actual level of DNA methylation, in particular on a single nucleotide resolution in genomic DNA. 
\vspace{0.3cm}
\\
At some point, this research indicated that  the obtained DNA methylation level\footnote{e.g., \cite{yu} refers to it as "total DNA methylation"} can be split into  hydroxymethylcytosine (5hmC) and 5-methylcytosine (5mC) components, with 5mC playing an important role in gene silencing and genome stability \cite{Booth1}.  The  second component, 5hmC methylation, was first discovered in 2009  as an another  form of cytosine modification \cite{yu}. Since then, its function as an intermediate in active DNA demethylation and an important  epigenetic regulator of mammalian  development,  as well as   its role as a possible epigenetic mark impacting genome stability  has come into  the spotlight \cite{Booth1, godderis, nazor, bachman}. At that point, the question concerning  reliable detection and accurate quantification of  5hmC emerged. 
\vspace{0.3cm}
\\
Until now, two key techniques  for the quantification of 5hmC levels, the TET-assisted bisulfite sequencing ({\it TAB-seq}) technique and the oxidative bisulfite sequencing ({\it oxBS-seq}) technique\footnote{An alternative method that can also be applied for the simultaneous quantification of 5mC and 5hmC, the so-called {\it liquid chromatography-tandem mass spectrometry (LC-MS/MS) method}, is presented in \cite{godderis, bachman}. In particular,  \cite{bachman} shows that 5hmC is an oxidation product of 5-methylcytosine which arises slowly within the first 30 hours after DNA synthesis and remains (almost) constant during the cell cycle.}\footnote{Our research results are based on the real data derived by means of the {\it oxBS}-technique.}, were established. When applied for the quantification of 5hmC methylation, the  {\it TAB-seq} technique uses the fact that 5mC can be converted to 5hmC in mammalian DNA by TET emzymes \cite{yu, nazor}. In the context of this technique, 5hmC sites are blocked by means of  $\beta$-glucosyltransferase ($\beta$GT) in the first step. Then a recombinant mouse TET1 enzyme is applied to convert 5mC to 5caC. Finally, by means of bisulfite treatment, 5caC is converted to uracil, leaving only glucosylated 5hmC to be read as a cytosine. Note that  the {\it TAB-seq} technique is known to be cost-intensive due to the use of TET1 protein \cite{Booth1}; this may become an issue when applying this technique for 5hmC quantification.
\vspace{0.3cm}
\\
In the  context of  the second technique, DNA methylation levels can be obtained from the bisulfite sequencing ({\it BS-seq}) procedure \cite{Booth1}. However, this procedure can only differentiate between methylated and unmethylatedd cytosine bases, and cannot discriminate between  5mC and 5hmC.  
To determine the level  of 5hmC at a considered nucleotide position, the oxidative bisulfite sequencing ({\it oxBS-seq}) approach can be applied. This approach yields C's only at 5mC sites while oxidating 5hmC to 5-formylcytosine (5fC) and later converting them to uracil. As a result, an amount of 5hmC at each particular nucleotide position
can be determined as the difference between the {\it oxBS-seq}  (which identifies 5mC) and the {\it BS-seq} (which identifies 5mC+5hmC) readouts. Indeed, such substraction seems to make sense biochemically, even if from a statistical point of view it may clearly increase  the noise level in the assay. 
\vspace{0.3cm}
\\
In order to quantify the level of 5hmC in the context of the {\it oxBS}-technique,  the following quantity   is introduced in \cite{stewart, field}
\begin{eqnarray}\label{b12}
\Delta \beta^{oxBS} = \beta_{BS} - \beta_{oxBS}=\frac{M_{BS}}{M_{BS}+U_{BS}+100} - \frac{M_{oxBS}}{M_{oxBS}+U_{oxBS}+100}. 
\end{eqnarray}
Here $M$ is the intensity of the methylated allele, $U$ is the intensity of the unmethylated allele, $\beta_{BS}$ is the methylation level obtained from the {\it BS-seq} method, and $\beta_{oxBS}$ is the methylation level derived by means of the {\it oxBS-seq} method. As stated in \cite{stewart, field}, the quantity $\Delta \beta^{oxBS} $ has  to be  computed for each single CpG and each single sample\footnote{Symmetrically, in cases where hydroxymethylation is to be quantified in the context of the {\it TAB-seq -} method, the quantity 
\begin{eqnarray}
\Delta \beta^{TAB} = \beta_{BS} - \beta_{5hmC}=\frac{M_{BS}}{M_{BS}+U_{BS}+100} - \frac{M_{5hmC}}{M_{5hmC}+U_{5hmC}+100}, 
\end{eqnarray} 
with $\beta_{BS}$ as above and $\beta_{5hmC}$ derived by means of  the TAB-seq method, can be computed for each single probe and each single CpG; for more details see \cite{nazor}.}  and can be interpreted  as a \enquote{measure of hydroxymethylation} and \enquote{a reflection of the 5hmC level at each particular probe location} \cite{stewart}. This measure can then be applied in the screening step so as to exclude from  further analysis those CpGs that do not appear to contain hydroxymethylation.  
\vspace{0.3cm}
\\
Due to its definition, $\Delta \beta^{oxBS}$  can take values between -1 and 1, where negative values of $\Delta \beta^{oxBS}$  \enquote{represent false differences in methylation score between paired BS-only and oxBS data sets} and may be interpreted as a \enquote{background noise} \cite{stewart}. This interpretation has meanwhile been questioned in \cite{housemann}, where the authors discuss the \enquote{naive} estimation of the 5hmC level via the difference of two $\beta$ values as proposed in  \cite{stewart, field, nazor} and introduce a model for describing and estimating the proportions of 5mC and 5hmC via beta distributed random variables. The aim of such modeling was to  disallow negative proportions; the corresponding model is implemented in the the $R$-package {\it OxyBS}.
\vspace{0.3cm}
\\ 
While using $\Delta \beta^{oxBS}$ for the identification  of  significantly hydroxymethylated cytosines, the issue of  an appropriate {\it $\Delta \beta^{oxBS}$ threshold} arises.  In \cite{field}, the authors introduced  $\Delta \beta^{oxBS} > 0$ as an indicator for a given CpG to be {\it hydroxymethylated}; for all CpGs with low or negligible levels of 5hmC, $\Delta \beta^{oxBS} $ \enquote{may be negative as a consequence of inevitable random noise.}
In \cite{stewart}, the  threshold for $\Delta \beta^{oxBS}$  has been set to $0.3$ or $30 \%.$ 
However, it is not evident, whether the thresholds for $\Delta \beta^{oxBS}$ proposed in \cite{stewart, field} can be applied for any given data set or whether such threshold should be derived for each data set separately.  
\vspace{0.28cm}
\\
In the present paper we will first address the applicability of the $\Delta \beta^{oxBS}$ measure (in the following notation just $\Delta \beta$) for quantification of 5hmC levels and indicate limitations of this measure. Therefore we will discuss properties of $\Delta \beta$, both analytically and on data. Then we will introduce a number of alternative hydroxymethylation measures and compare their properties and similarity with those of $\Delta \beta$.  All results will be illustrated by means of real data analyses. 
\vspace{0.3cm}
\\
Data analyses presented here were performed on two  real data sets derived from  brain and whole blood tissues. This fact makes these analyses particularly interesting, since global 5hmC levels  are known to differ substantially between different tissue types  \cite{stewart} and, in particular, human brain is known to have the highest global levels of both 5hmC and 5mC, with more than 1000 times greater than the levels in blood \cite{Booth1,nestor,laird}.   Note that  also in  \cite{stewart} the $\Delta \beta$ measure is analyzed on brain and whole blood tissue.

%%%%%%%%%%%%%%%%%%%%%%%%%%%%%%%%%%%%%%%%%%%%%%%%%%%%%%%%%%%%%%%%%%%%%%%%%%
%%%%%%%%%%%%%%%%%%%%%%%%%%%%%%%%%%%%%%%%%%%%%%%%%%%%%%%%%%%%%%%%%%%%%%%%%%

%%%%%%%%%%%%%%%%%%%%%%%%%%%%%%%%%%%%%%%%%%%%%%%%%%%%%%%%%%%%%%%%%%%%%%%%%%
%%%%%%%%%%%%%%%%%%%%%%%%%%%%%%%%%%%%%%%%%%%%%%%%%%%%%%%%%%%%%%%%%%%%%%%%%%

%%%%%%%%%%%%%%%%%%%%%%%%%%%%%%%%%%%%%%%%%%%%%%%%%%%%%%%%%%%%%%%%%%%%%%%%%%

\section{On the applicability of $\Delta \beta$ as a measure  for 5hmC  levels}\label{sec1}
Given the methylated and unmethylated  intensities $M$ and $U,$ the methylation level of the particular probe  can be described by  the {\it methylation proportion}
\begin{eqnarray}\label{beta}
\beta =  \frac{M}{M+U+ 100}
\end{eqnarray}
as introduced in  \cite{bib}.
Thus, the 5hmC measure $\Delta \beta^{oxBS}$  in (\ref{b12}) is just the difference of two methylation proportions as derived from $BS$-$seq$ and $oxBS$-$seq$ treatment, respectively. However, this simple definition, while appearing to be plausible at first,  leads to a number of ambiguities. First, the outcomes of $\Delta \beta$ are usually interpreted as follows \cite{stewart}:  Positive values of $\Delta \beta $ are taken as an indicator for substantial hydroxymethylation, whereas  small values of  $\Delta \beta $ indicate no or only nonsubstantial hydroxymethylation.  Negative values of $\Delta \beta$  are considered as resulting from  background noise. In the sequel, we will analyze each of these cases individually and show the  limitations of  the above-mentioned interpretations. In addition, we will have a closer look at the correction term 100 appearing in the denominators in (\ref{b12}) and address possible consequences of this particular  choice.
\vspace{0.3cm}
\\
The first ambiguity arising from (\ref{b12}) concerns the application of $\Delta \beta$ as a 5hmC measure in general. Even if both components in the difference (\ref{b12}) do represent the respective methylation proportions for {\it BS} and {\it oxBS} data, these proportions are  calculated on two different  bases: the proportion $\beta_{BS}$ represents the methylation proportion based on the  global {\it BS} methylation intensity $M_{BS}+U_{BS}$, whereas the proportion $\beta_{oxBS}$ represents the methylation proportion based  on the global {\it oxBS} methylation intensity $M_{oxBS}+U_{oxBS}$. Thus, a direct comparison of these two proportions  is difficult to justify and, as a result, the interpretation of $\Delta \beta$ as \enquote{a reflection  of the 5hmC level at each particular probe} suggested in \cite{stewart} is not well founded. A graphical illustration of this issue is presented in Figure \ref{ss1}. In particular, as that figure shows, all ten simulated data points satisfy both   $$M_{BS}<M_{oxBS}\,\,\, \text{and}\,\,\, U_{BS} < U_{oxBS} $$
simultaneously. 

\begin{center}
\begin{figure}[t!]
\begin{overpic}[width=16cm]{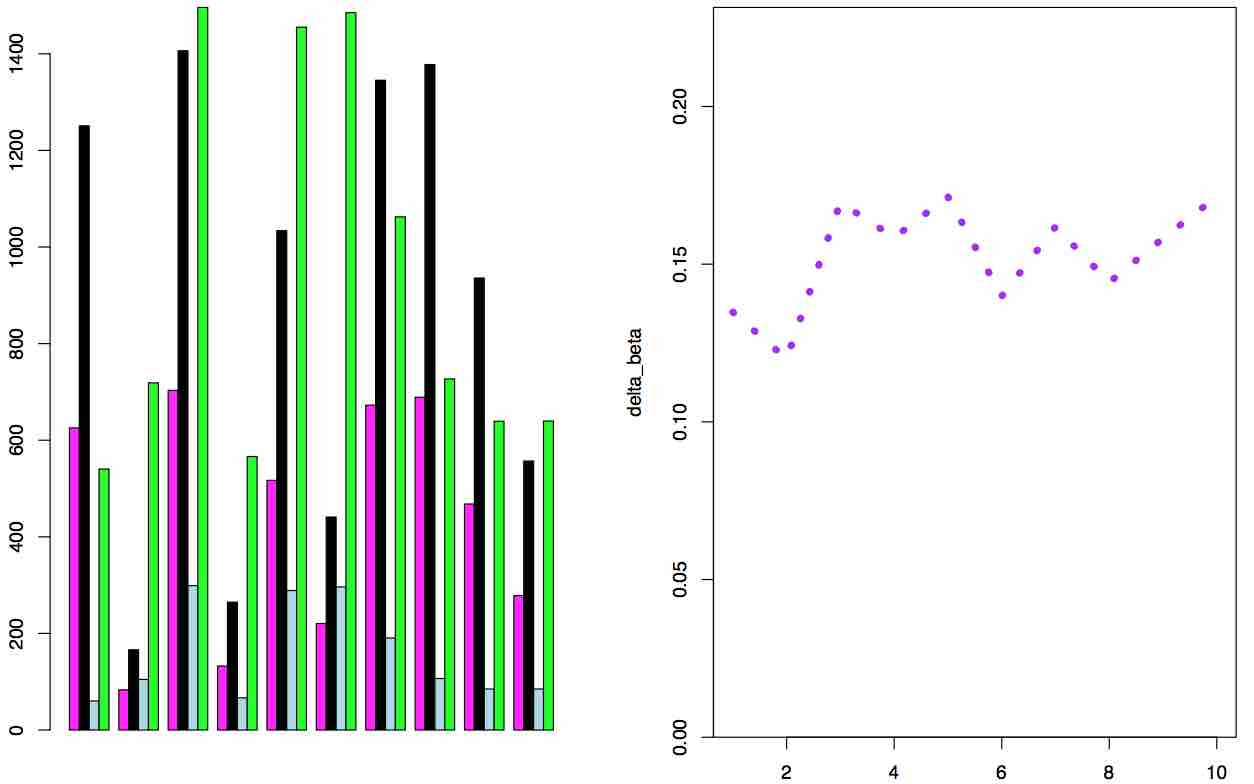}
\put(27,0){\scriptsize{s1}}
\put(46,0){\scriptsize{s2}}
\put(63,0){\scriptsize{s3}}
\put(82,0){\scriptsize{s4}}
\put(100,0){\scriptsize{s5}}
\put(118,0){\scriptsize{s6}}
\put(136,0){\scriptsize{s7}}
\put(153,0){\scriptsize{s8}}
\put(171,0){\scriptsize{s9}}
\put(187,0){\scriptsize{s10}}
\put(17,292){intensities}
\put(257,292){$\Delta \beta$}
 \end{overpic}
  \caption{On the applicability of $\Delta \beta$  as a 5hmC measure: the case with $\Delta \beta>0.$ The magenta bars correspond to the $M_{BS}$ intensities,  the black bars describe the $M_{oxBS}$ intensities, the blue bars correspond to the $U_{BS}$ intensities, and the green bars correspond to the $U_{oxBS}$ intensities. It is clear from the left-hand panel that none of the data points show any substantial 5hmC level, but due to   $\Delta \beta>0$ they will nevertheless be selected as hydroxymethylated probes. \vspace{0.5cm}}
 \label{ss1}
 \end{figure}
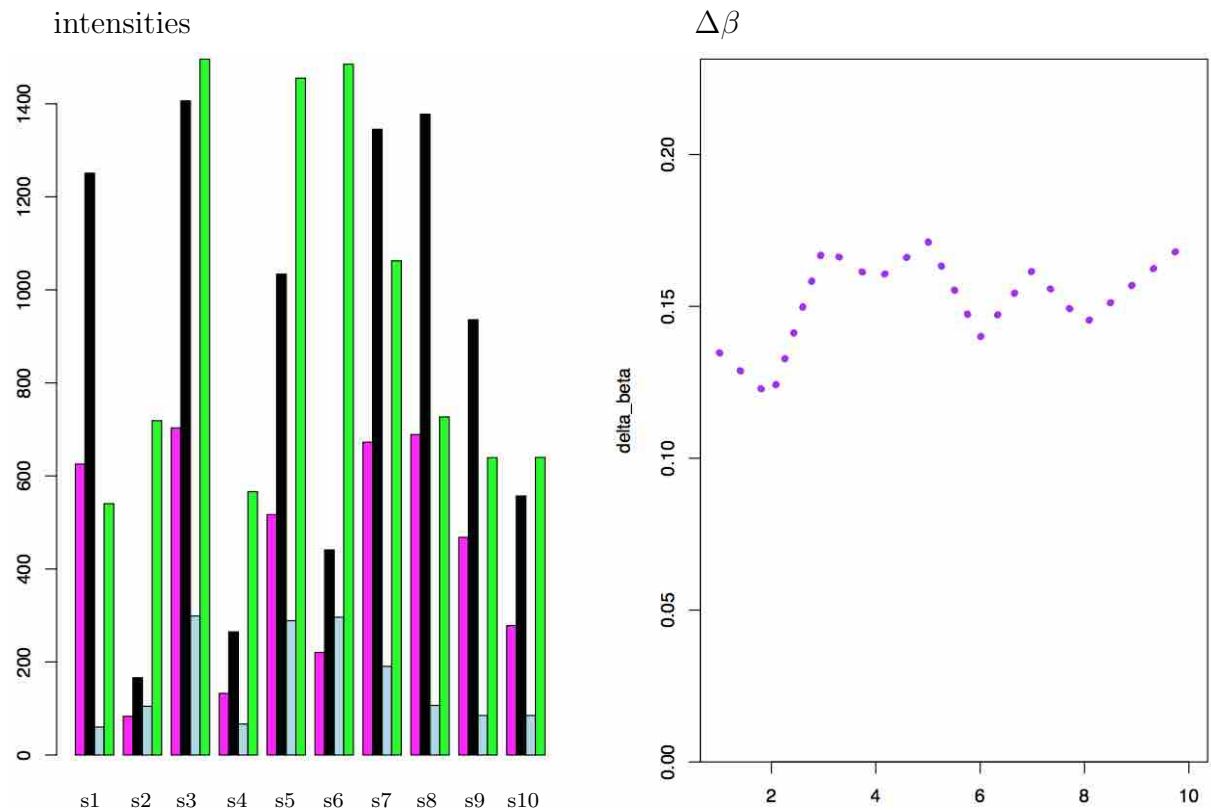
\end{center}
\vspace{-1cm}
That is, for each of these ten points the {\it BS} intensities are lower than the {\it oxBS}  intensities which intuitively corresponds to the interpretation  "no positive 5hmC observed". On the other hand, the condition $\Delta \beta>0$ holds for each of ten considered data points.
\vspace{0.3cm}
\\
\noindent Another ambiguity arising in  using the measure $\Delta \beta$ is an adequate  interpretation of its negative outcomes. In \cite{stewart}, the authors state that only probes with $\Delta \beta >0$ \enquote{represent potential sites of 5hmC} and that negative values of $\Delta \beta$ \enquote{...are likely to reflect background noise generated by the method...}; this view was also shared in \cite{field}. However, such an interpretation does not seem to be plausible according to our discussion on $\Delta \beta$ as a difference of two methylation proportions  $\beta_{BS}$ and  $\beta_{oxBS}$. Figure \ref{ss2} illustrates  this issue for ten simulated data points that satisfy both 
$$M_{BS}>M_{oxBS}\,\,\, \text{and}\,\,\, U_{BS} > U_{oxBS}, $$ although  the condition $\Delta \beta <0$ holds. Thus, these data points show a positive 5hmC level due to their  {\it BS} intensities exceeding their  {\it oxBS} intensities, but they will not be detected by  the measure $\Delta \beta $ as being substantially hydroxymethylated. 
\begin{center}
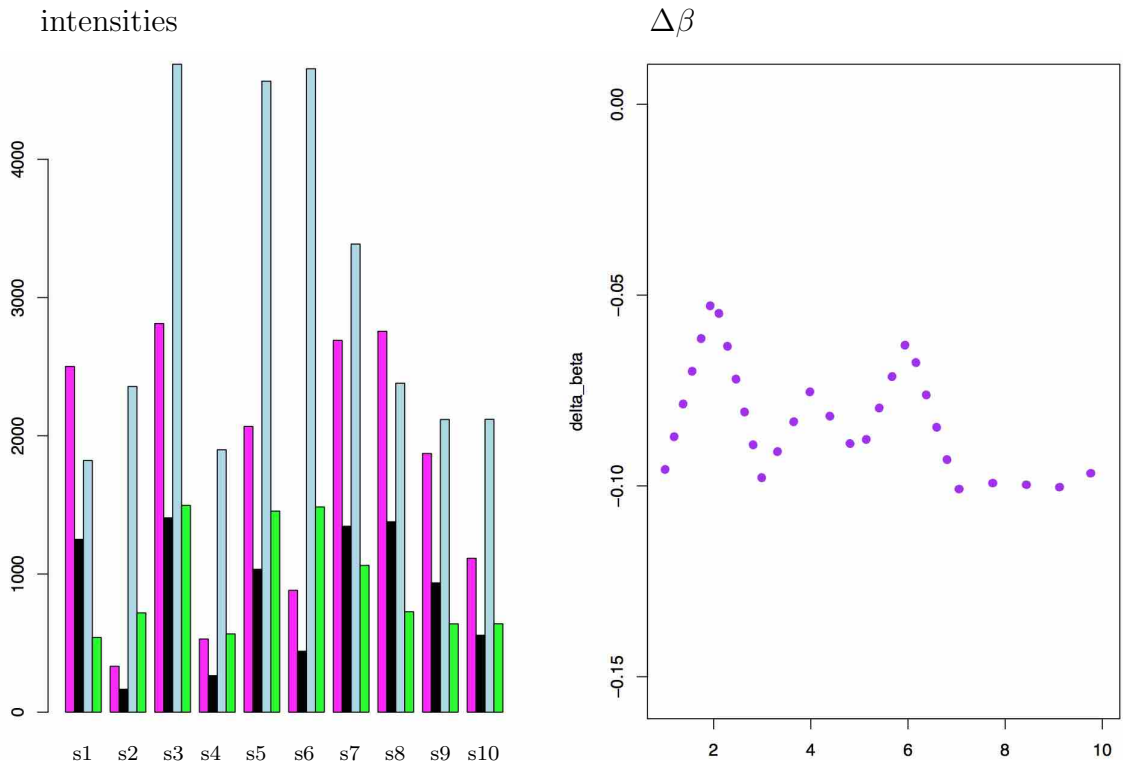
\begin{figure}[t!]
\begin{overpic}[width=15cm]{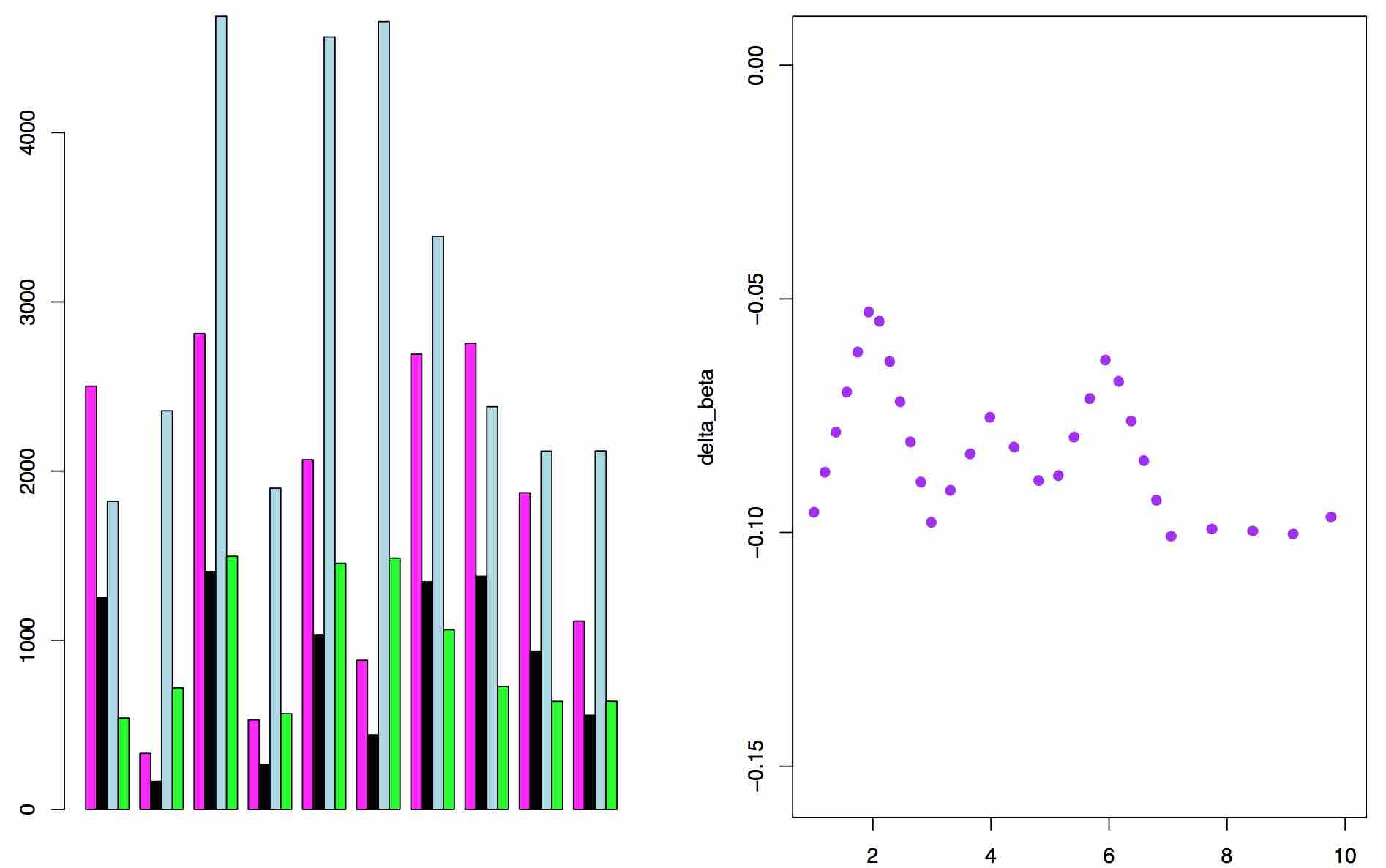}
\put(29,0){\scriptsize{s1}}
\put(46,0){\scriptsize{s2}}
\put(63,0){\scriptsize{s3}}
\put(77,0){\scriptsize{s4}}
\put(94,0){\scriptsize{s5}}
\put(112,0){\scriptsize{s6}}
\put(129,0){\scriptsize{s7}}
\put(146,0){\scriptsize{s8}}
\put(163,0){\scriptsize{s9}}
\put(177,0){\scriptsize{s10}}
\put(17,275){intensities}
\put(245,275){$\Delta \beta$}
 \end{overpic}

        \caption{On the applicability of $\Delta \beta $  as a 5hmC measure: the case with $\Delta \beta<0.$ Notations are the same as in Figure \ref{ss1}. Note that the left-hand panel suggests that all ten data points do exhibit a substantial level of 5hmC, whereas the right-hand panel shows negative $\Delta \beta$ values.\vspace{0.4cm} }
        \label{ss2}
\end{figure}
\end{center}
\vspace{-0.88cm}
\noindent 
One of the main advantages of the measure  $\beta$, which has also contributed to its common application  as a methylation measure, is its intuitive interpretation as an approximation of the {\it percentage of methylation} \cite{pan}; here $\beta=0$ denotes unmethylated probes and $\beta=1$ denotes fully methylated probes. Unfortunately, this interpretation does not carry over to the measure $\Delta \beta$. Indeed, in (\ref{b12}) the condition $\Delta \beta =0$ solely implies   $$\frac{M_{BS}}{M_{oxBS}} = \frac{U_{BS}+100}{U_{oxBS}+100}$$ and it is unclear how this last equality should be interpreted in terms of the observed 5hmC level.  Moreover, Figure  \ref{ss3} demonstrates that we  can get $\Delta \beta=0$ in cases where $$M_{BS}<M_{oxBS}\,\,\, \text{and}\,\,\, U_{BS} < U_{oxBS}$$ (i.e., "no substantial 5hmC level observed", see both upper panels in Figure  \ref{ss3}) as well as in cases where $$M_{BS}>M_{oxBS}\,\,\, \text{and}\,\,\, U_{BS} > U_{oxBS}$$ (i.e., "a substantial 5hmC level observed", see both lower panels in Figure  \ref{ss3}).

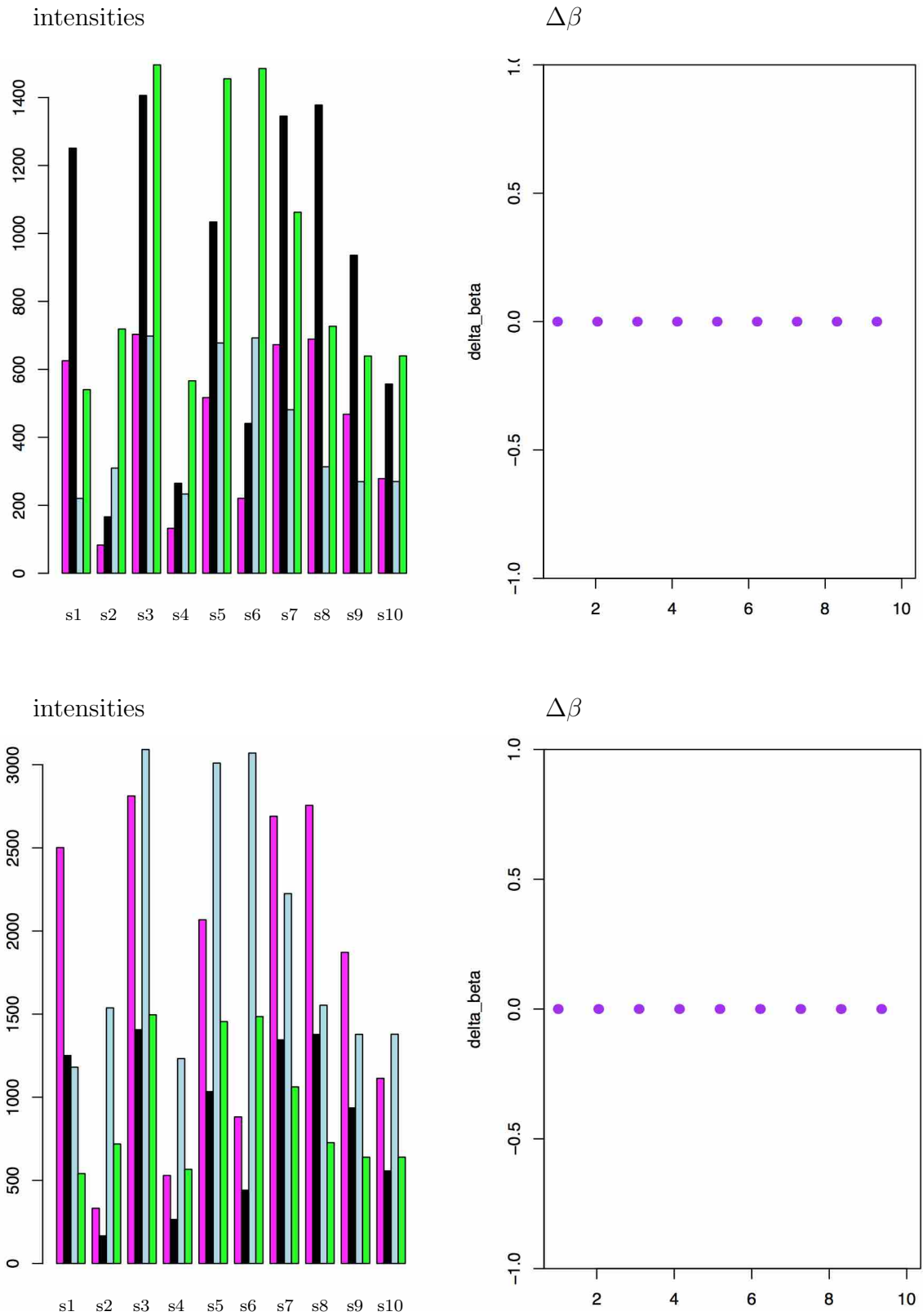
\begin{figure}[htp!]
\centering
\begin{minipage}{0.9\textwidth}
%\begin{figure}
\begin{overpic}[width=15cm]{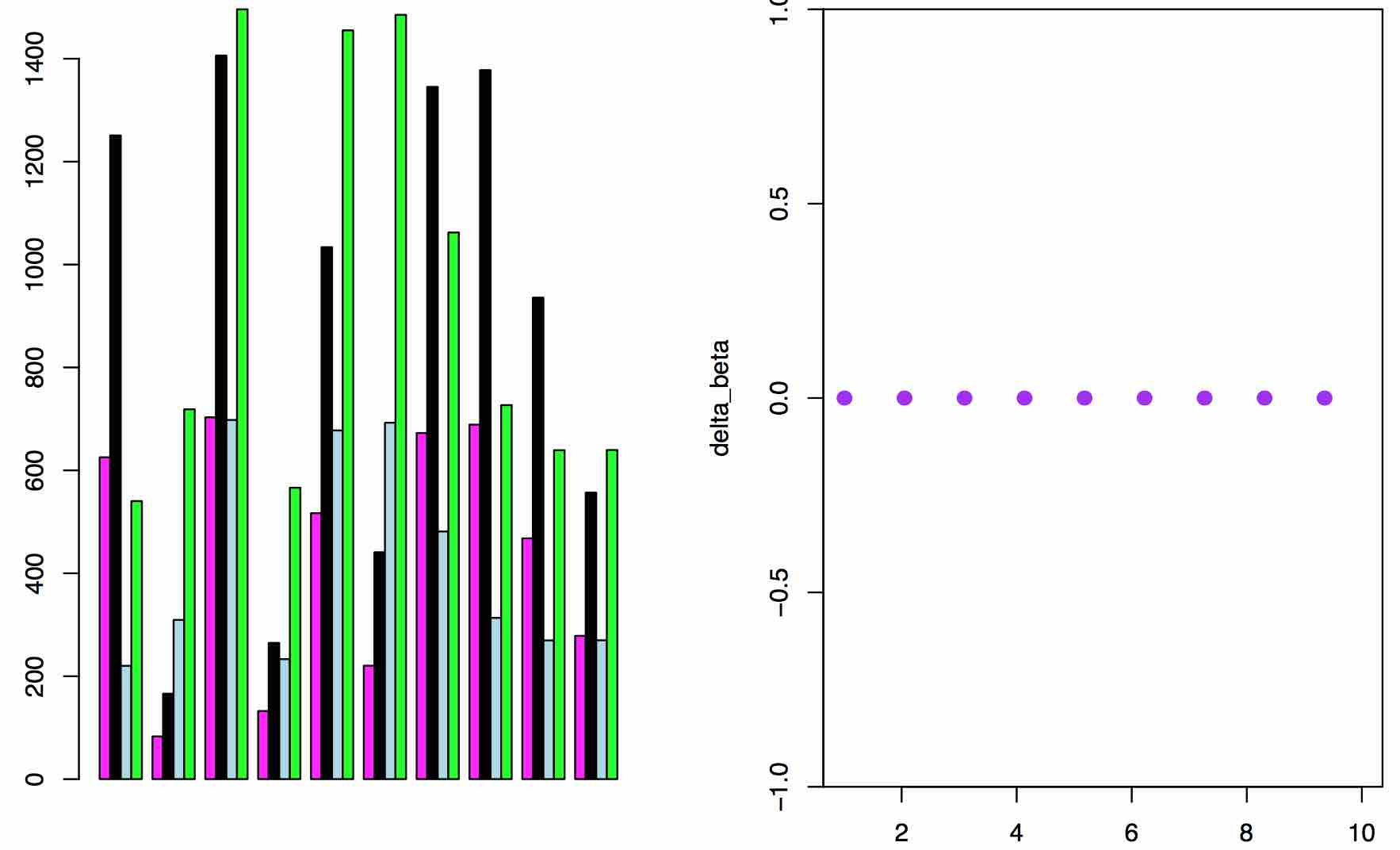}
\put(32,0){\scriptsize{s1}}
\put(48,0){\scriptsize{s2}}
\put(65,0){\scriptsize{s3}}
\put(81,0){\scriptsize{s4}}
\put(98,0){\scriptsize{s5}}
\put(114,0){\scriptsize{s6}}
\put(131,0){\scriptsize{s7}}
\put(146,0){\scriptsize{s8}}
\put(161,0){\scriptsize{s9}}
\put(175,0){\scriptsize{s10}}
\put(17,275){intensities}
\put(252,275){$\Delta \beta$}
 \end{overpic}
 %\end{figure}
 \end{minipage}
\\
\vspace{1.85cm}
\begin{minipage}{0.9\textwidth}
\centering
% \begin{figure}
\begin{overpic}[width=15cm]{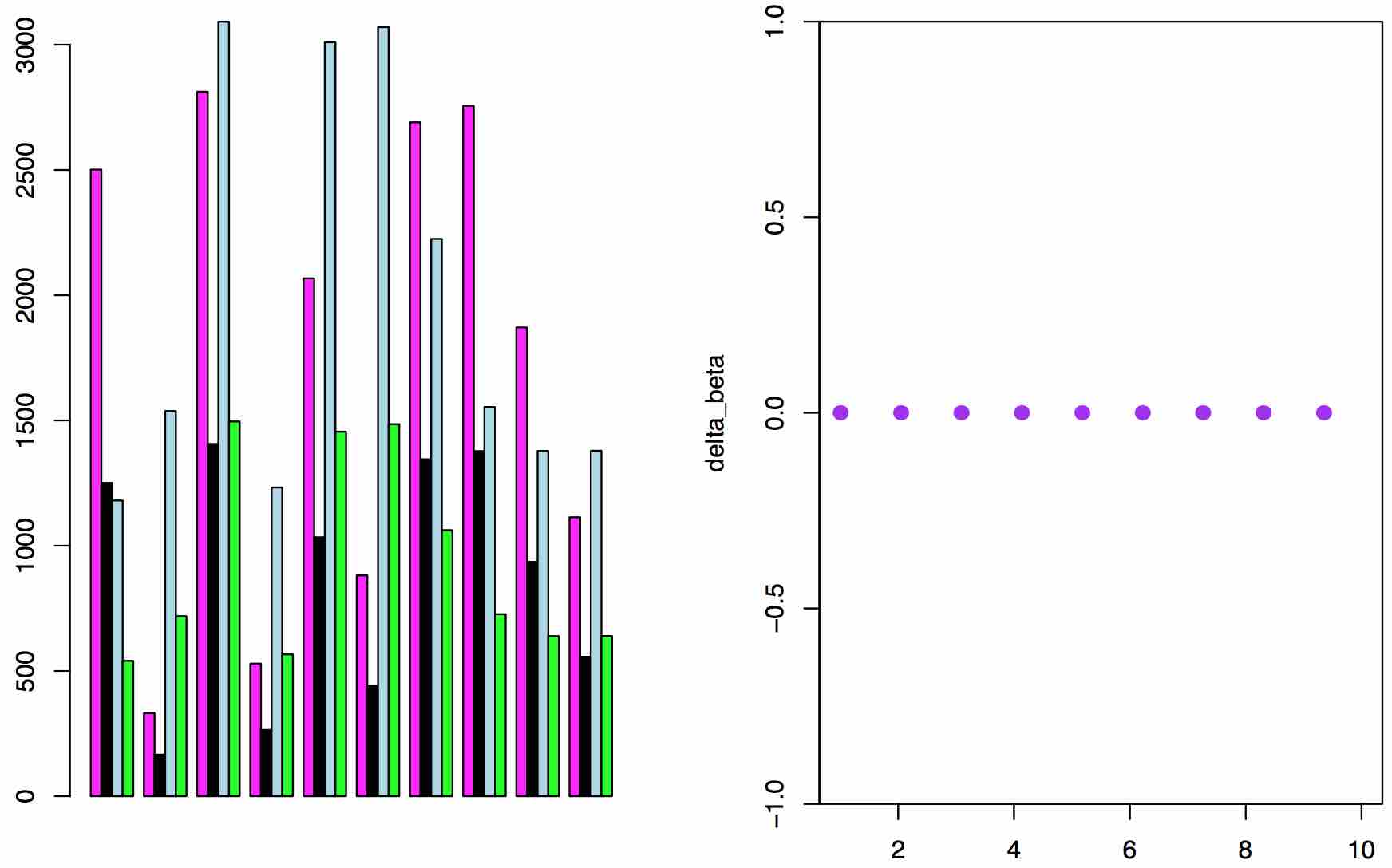}
\put(29,0){\scriptsize{s1}}
\put(46,0){\scriptsize{s2}}
\put(63,0){\scriptsize{s3}}
\put(79,0){\scriptsize{s4}}
\put(97,0){\scriptsize{s5}}
\put(112,0){\scriptsize{s6}}
\put(129,0){\scriptsize{s7}}
\put(145,0){\scriptsize{s8}}
\put(161,0){\scriptsize{s9}}
\put(175,0){\scriptsize{s10}}
\put(17,275){intensities}
\put(252,275){$\Delta \beta$}
 \end{overpic}

     \caption{On the applicability of $\Delta \beta $  as a 5hmC measure: the case with $\Delta \beta=0.$ Notations are the same as in Figure \ref{ss1} above.}
        \label{ss3}
 %\end{figure}
\end{minipage}
\end{figure}

\vspace{0.3cm}
\noindent Altogether, our analyses of the three cases $\Delta\beta>0$, $\Delta\beta=0$, and $\Delta\beta<0$ has shown that their usual respective interpretations as indicators of substantial hydroxymethylation, no hydroxymethylation, and noise  are problematic.

\begin{figure}[h!]
             \hspace{0.75cm} \includegraphics[width=0.9\linewidth, height=0.35\textheight]{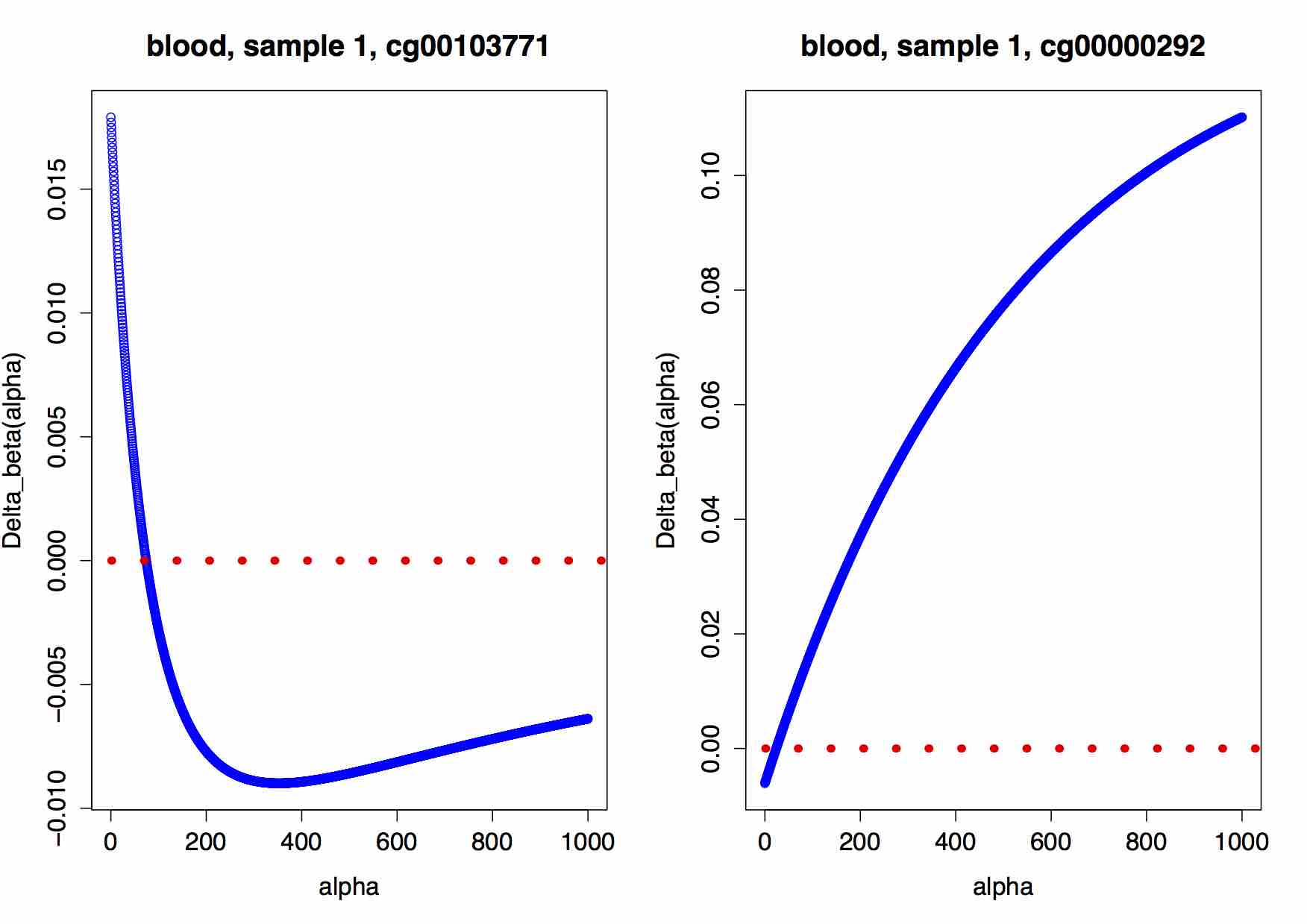}
        \caption{Dependence of $\Delta \beta(\alpha) $  on the choice of $\alpha$: $\Delta \beta(\alpha) $ changing its sign from positive to negative  (left-hand panel) and from negative to positive  (right-hand panel) as $\alpha$ increases. Note that this result refers to a given CpG site and a given sample. \vspace{0.75cm}}
        \label{ns0}
\end{figure}

\vspace{0.3cm}
\noindent Next, when analyzing the 5hmC measure $\Delta \beta$ in  (\ref{b12}), the question arises  why one chooses the number 100 in the denominators $M_{BS}+U_{BS}+100$ and $M_{oxBS}+U_{oxBS}+100$. This choice  seems to stem from the practical convention in the definition of $\beta$-values \cite{pan}, and thus was carried over to the definition of $\Delta\beta$ as well  \cite{stewart, nazor}. As a matter of fact, there is no strong reason why  the correction term 100 in (\ref{beta}) should not be replaced with another value $\alpha>0$. This leads to the following more general  definition of the methylation proportion
\begin{eqnarray}\label{uu}
\beta(\alpha) =  \frac{M}{M+U+ \alpha},
\end{eqnarray} 
with $ \alpha > 0$.

\vspace{0.3cm}
\noindent While one can safely argue that the actual choice of the parameter $\alpha$ is not crucial for the interpretation of the methylation proportion $\beta(\alpha)$ itself (see \cite{pan}), we will now argue that the choice of $\alpha$ can become critical when using the sign of the quantity  
\begin{eqnarray}\label{b112}
\Delta \beta(\alpha) = \beta_{BS} - \beta_{oxBS}=\frac{M_{BS}}{M_{BS}+U_{BS}+\alpha} - \frac{M_{oxBS}}{M_{oxBS}+U_{oxBS}+\alpha},
\end{eqnarray}
as an indicator for  hydroxymethylation.  To this end, we will show in Appendix \ref{a2} that, under certain conditions, the sign of  $\Delta\beta(\alpha)$ can change  from positive to negative or vice versa if $\alpha$ varies; for an illustration see Figure \ref{ns0}.
 We will also give a formula for the corresponding point of sign change, which we   henceforth denote by $\alpha^*$. 
 \vspace{0.3cm}
 \\
 \noindent In Table 1, we show that a notable percentage of CpGs in our data sets of blood and brain tissue exhibits a sign change of $\Delta\beta(\alpha)$. Of these, a substantial percentage has a value of $\alpha^*$ being less than or equal to 1000. From a practical point of view, these  results imply that stating whether or not a particular CpG exhibits a  positive  level of 5hmC can depend strongly on the choice of the correction parameter $\alpha$.

\vspace{0.2cm}
\begin{figure}[h!]
\includegraphics[width=1\linewidth, height=0.165\textheight]{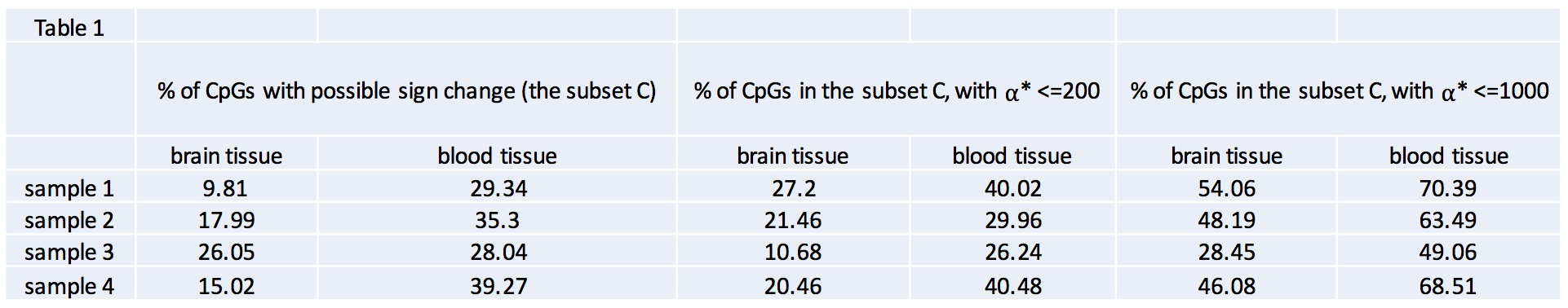}

\vspace{0.3cm}
Table 1: This table demonstrates that for a substantial percentage of CpGs in our data sample the measure $\Delta \beta(\alpha)$ may  change its sign for varying $\alpha$. Since a noticeable part of such CpGs change their sign in a given point $\alpha^*$  left from 200, we can conclude that even small deviations from the chosen value $\alpha=100$  may lead to considerable changes in the set of CpGs that are flagged as exhibiting a positive value of 5hmC. In this sense, the measure $\Delta\beta(\alpha)$ is not robust with respect to the choice of the correction parameter $\alpha$.  \vspace{0.35cm}
       
\end{figure}

\vspace{1cm}
\begin{figure}[htp!]
    \centering
    \vspace{-0.35cm}
\begin{overpic}[width=12cm]{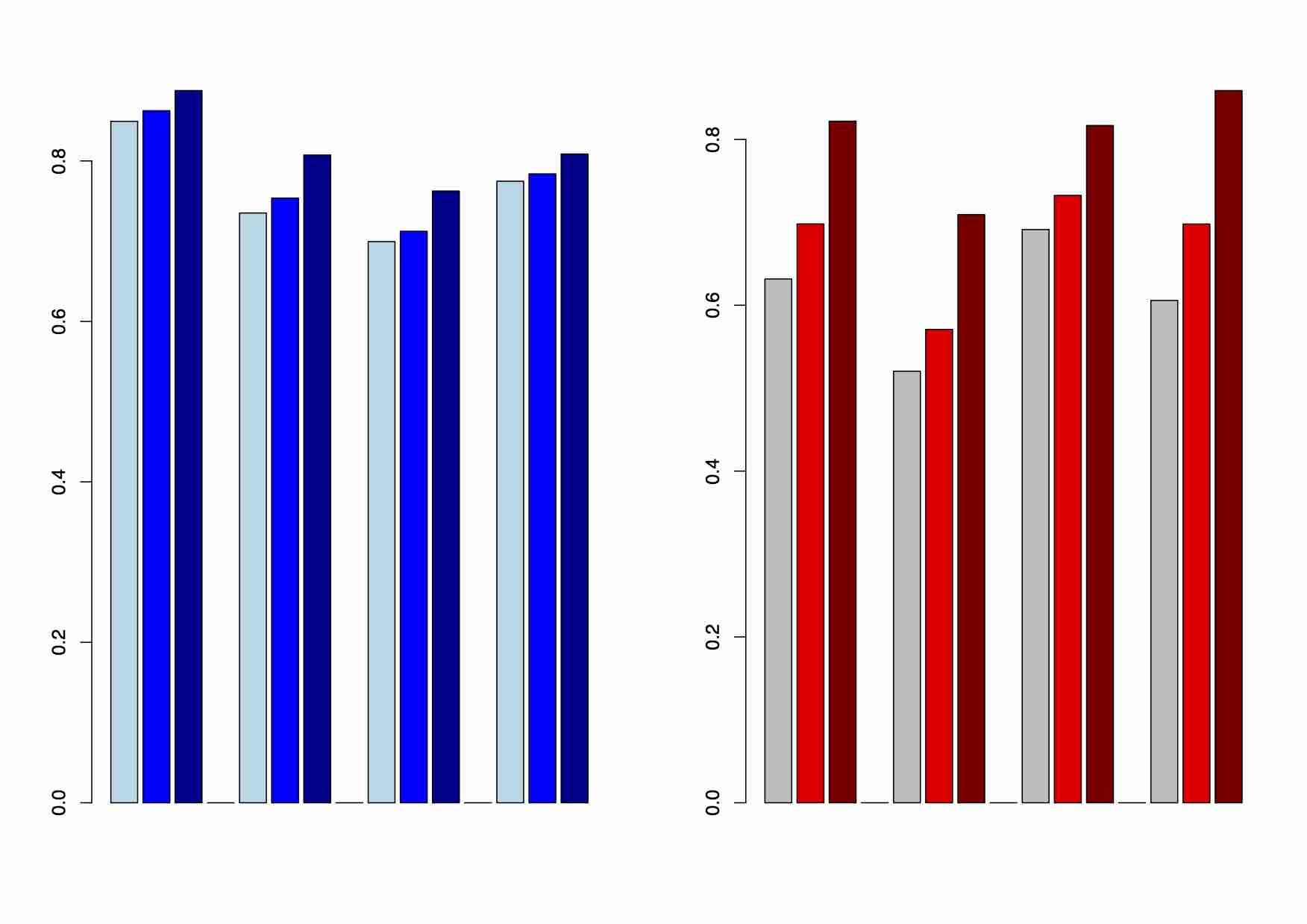}
\put(36,10){s1}
\put(72,10){s2}
\put(103,10){s3}
\put(140,10){s4}
\put(210,10){s1}
\put(240,10){s2}
\put(277,10){s3}
\put(308,10){s4}
\put(25,235){brain tissue}
\put(205,235){blood tissue}
 \end{overpic}
\caption{Percentage of CpGs  satisfying $\Delta \beta(\alpha) > 0$: a real data example for four given samples and both given tissues.  Here $\alpha=10$ (light blue bars for brain and grey bars for blood tissue), $\alpha=100$ (blue bars for brain and red bars for blood tissue)  and $\alpha=1000$ (dark blue bars  for brain and dark red bars for blood tissue) were considered. Note that the considered percentage increases with increasing values of $\alpha$ and that this effect is most observable in case of blood tissue. \vspace{0.35cm}}
        \label{f055}

 %   \end{minipage}
\end{figure}

\vspace{0.1cm}
\noindent In the context of the dependence of $\Delta \beta(\alpha)$ on  the choice of  $\alpha $, a question concerning the possible impact of  this choice on the percentage of CpG sites  satisfying the condition $ \Delta \beta(\alpha)>0$ arises, for each given sample. As Figure \ref{f055} suggests, this percentage converges to a certain constant value as  $\alpha$ increases. For more details on this topic see the discussion below.

\vspace{0.4cm}
\noindent Finally, it  follows from (\ref{b112}) that $\Delta \beta$   ranges between -1 and 1. Such a limited range of possible values  can become an issue, for instance,  if after the completion of the screening step  one is looking  for appropriate statistical methods for the further analysis of the $\Delta \beta$ values, such as linear regression.
\vspace{0.3cm}
\\
To summarize, in the present section we stated a couple of limitations of the 5hmC measure $\Delta \beta(\alpha)$ which make its practical applicability for quantification of 5hmC levels questionable. In the next section we will address these limitations and propose a number of alternative 5hmC measures.
%%%%%%%%%%%%%%%%%%%%%%%%%%%%%%%%
%%%%%%%%%%%%%%%%%%%%%%%%%%%%%%%%

%%%%%%%%%%%%%%%%%%%%%%%%%%%%%%%%%%%%%%%%%%%%%%%%%%%%%%%%%%%%%%%%%%%%%%%
%%%%%%%%%%%%%%%%%%%%%%%%%%%%%%%%%%%%%%%%%%%%%%%%%%%%%%%%%%%%%%%%%%%%%%%

 %%%%%%%%%%%%%%%%%%%%%%%%%%%%%%%%%%%%%%%%%%%%%%%%%%%%%%%%%%%%%%%%%%%%%%%
 %%%%%%%%%%%%%%%%%%%%%%%%%%%%%%%%%%%%%%%%%%%%%%%%%%%%%%%%%%%%%%%%%%%%%%%
 
 \section{Alternative 5hmC measures}\label{sec22}
 
  As discussed in the previous section, the 5hmC measure $\Delta \beta(\alpha)$ demonstrates a number of shortcomings such as a lack of interpretation or its dependence on the choice of the correction term $\alpha$. To overcome such shortcomings, alternative 5hmC measures may become a solution. 
 \vspace{0.3cm}
 \\ 
 Thus, in the present section we will first introduce a number of alternative  measures which can be applied for detection of CpGs with a positive level of 5hmC. Further, following the analysis pattern proposed for the measure $\Delta \beta(\alpha)$,  we will discuss  the basic properties  of these measures and compare them to $\Delta \beta(\alpha)$.   

%%%%%%%%%%%%%%%%%%%%%%%%%%%%%%%%
%%%%%%%%%%%%%%%%%%%%%%%%%%%%%%%%
%%%%%%%%%%%%%%%%%%%%%%%%%%%%%%%%

\begin{figure}[htp]
    \hspace{1.1cm} \includegraphics[width=0.9\linewidth, height=0.35\textheight]{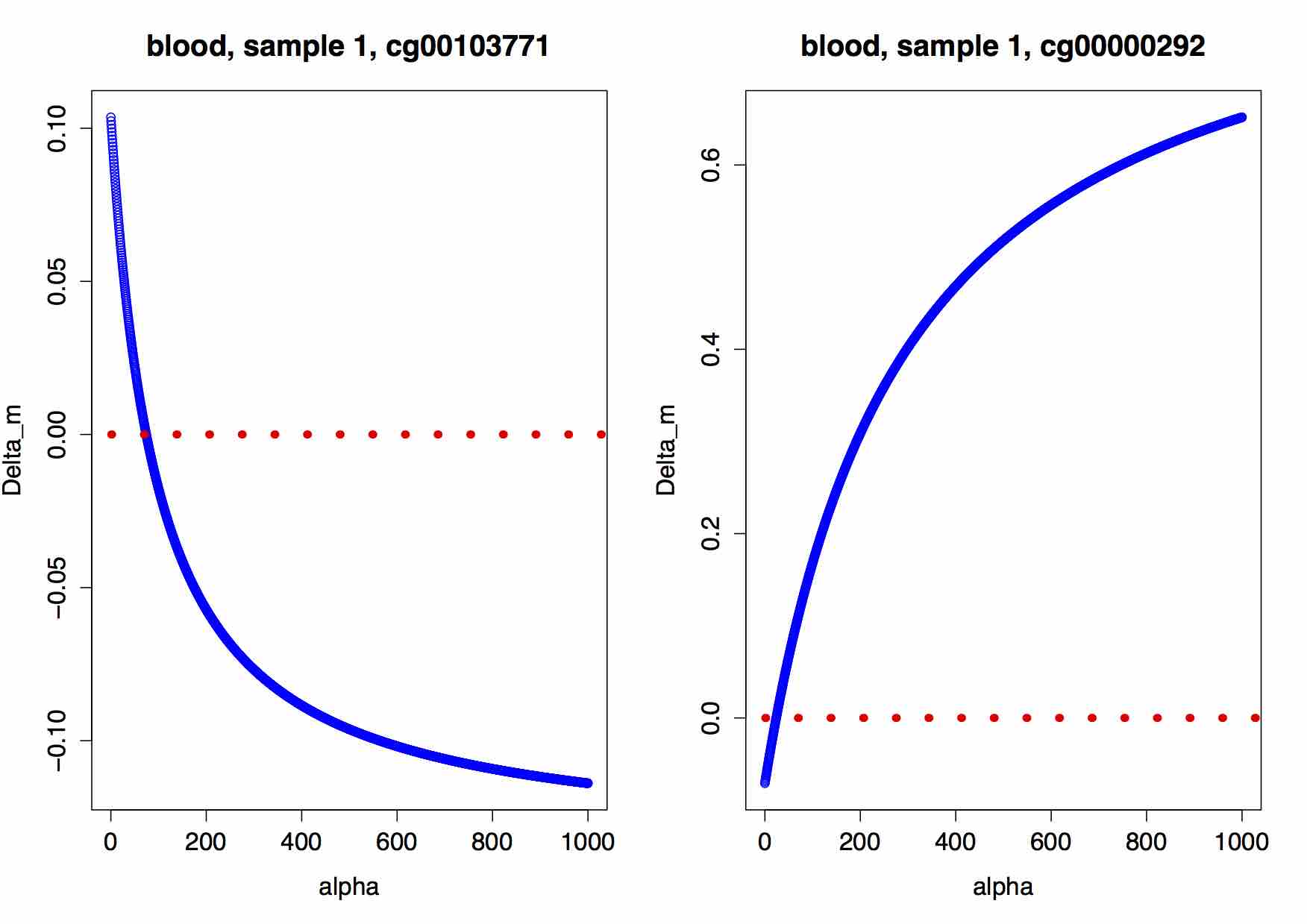}
        \caption{$\Delta m(\alpha) $ changing its values from positive to negative  (left-hand panel) and from negative to positive  (right-hand panel). \vspace{0.25cm}}
        \label{f11bn2}
 \end{figure}

\vspace{0.3cm}
\noindent  The first 5hmC measure we propose is based on the so-called {\it $m$-value} \cite{pan} given by a transformation of $\beta=\beta(\alpha)$  as 
 \begin{eqnarray}\label{mv}
 m = \log_2 \frac{\beta}{1-\beta}.
 \end{eqnarray} 
 While the $m$-value does not have an immediate and straightforward  interpretation such as the measure $\beta$, it was shown that it can outperform $\beta$ in quantifying the level of methylation, at least at  high and low methylation levels; see \cite{pan} for more discussion. Moreover, due to an unbounded range of possible values, a wider spectrum of statistical methods can be used for the analysis of  such $m$-values as compared to the number of methods applicable to $\beta$ values. 
 \vspace{0.3cm}
 \\
 By adopting the idea of $\Delta \beta$ defined as a difference of two respective $\beta$ values, we now consider the difference of two respective $m$-values and introduce  the measure
\begin{eqnarray}\label{eq1}
\Delta m = m_{BS} - m_{oxBS} =\log_2 \frac{\beta_{BS}}{1-\beta_{BS}} -\log_2\frac{\beta_{oxBS}}{1-\beta_{oxBS}}
\end{eqnarray}
as a possible alternative to the 5hmC measure $\Delta \beta$.  Note that, in contrast to \eqref{mv},  there is no formal transformation between $\Delta m$ and $\Delta \beta$ that would render $\Delta m$ as a function of $\Delta\beta$; see the Appendix  \ref{fgh77} for more details.
%%%%%%%%%%%%%%%%%%%%%%%%%%%%%
%%%%%%%%%%%%%%%%%%%%%%%%%%%%%
\vspace{0.3cm}
\\
As before, we now make the dependence of $\beta$  and $\Delta \beta$ on the correction parameter $\alpha$ explicit by writing $\beta(\alpha)$ and $\Delta\beta(\alpha)$. This dependence then carries over to $m$ and $\Delta m$, so that we will henceforth write $m(\alpha)$ and $\Delta m (\alpha)$. This latter dependence can be made explicit by using  standard calculations to transform (\ref{eq1})  into
\begin{eqnarray}\label{ui}
\Delta m(\alpha)=\log_2\frac{M_{BS}}{M_{oxBS}}+\log_2\frac{U_{oxBS} +\alpha}{U_{BS}+\alpha}. 
\end{eqnarray}
This is the expression for $\Delta m(\alpha)$ we will use in the further discussions.
\vspace{0.3cm}
\\
 Next, we recall that  CpGs with positive $\Delta \beta(\alpha) $ are typically considered as showing a substantial level of 5hmC;  see, e.g., \cite{stewart}. In the Appendix \ref{wd} we will show that  the condition $\Delta m(\alpha) > 0$ holds in the same cases as the condition  $ \Delta \beta(\alpha) >0$ is satisfied. Thus we can state that
\begin{quote}
 {\it at the end of the screening step, the two criteria $\Delta m >0 $ and $\Delta \beta >0$ will flag  the same CpG sites.}  
\end{quote}
In this sense,  both hydroxymethylation measures are comparable and can be used {\it interchangeably} in detecting  CpGs with a positive level of 5hmC. On the other hand, given that  $\Delta m(\alpha)$  can take values on the entire real line, a wider range of statistical methods can be applied for the further analysis of this hydroxymethylation measure.  
%%%%%%%%%%%%%%%%%%%%%%%%%%%%%
%%%%%%%%%%%%%%%%%%%%%%%%%%%%%
\vspace{0.3cm}
\\
As a matter of fact, the two measures $\Delta \beta(\alpha)$ and $\Delta m(\alpha)$ will always have the same sign, for any given sample and CpG. This result, while contributing to the comparability of two 5hmC measures, will at the same time  lead to similar  limitations for the interpretation of the measure $\Delta m(\alpha)$ as for $\Delta\beta(\alpha)$. First, $\Delta m(\alpha)$ will evidently exhibit the same ambiguities in the interpretation of its values as $\Delta \beta(\alpha)$ does; that is, it is not evident how the conditions $\Delta m(\alpha)>0$, $\Delta m(\alpha)=0$ and $\Delta m(\alpha)<0$ can be interpreted in terms of the 5hmC level  observed for a given CpG and sample. 
Second, the value of $\Delta m(\alpha)$ will  obviously depend on the choice of the correction term $\alpha$, just as the measure $\Delta \beta(\alpha)$ does. In particular,  $\Delta m(\alpha)$ may change its sign from positive to negative and vice versa under the same conditions the measure $\Delta \beta(\alpha)$ does. This issue is discussed in the Appendix  \ref{fgh} and \ref{a22}; for an illustration see Figure \ref{f11bn2}. 
\vspace{0.4cm}
\\
\vspace{-0.63cm}
%%%%%%%%%%%%%%%%%%%%%%%%%%%%%%%%%
%%%%%%%%%%%%%%%%%%%%%%%%%%%%%%%%%
\\
\vspace{-0.53cm}

\noindent As already indicated in case of $\Delta \beta(\alpha)$, the ability of the 5hmC measure  to change its sign can have unwanted results in the context of the screening step, where this measure  is the criterion for selecting  CpGs with a substantial amount of 5hmC. In particular, in certain cases  the condition $\Delta m(\alpha) > 0$ is just a matter of an appropriate choice of the correction term $\alpha$. 

\vspace{0.3cm}
\noindent To summarize the results of this section, we state that, on the one hand, the measure $\Delta m(\alpha)$ inherits  most properties of the measure $\Delta \beta(\alpha)$ which are relevant for the selection procedure  and thus  both measures can be used interchangeably while detecting the CpGs with a substantial level of 5hmC; however, $\Delta m(\alpha)$ still lacks an appropriate interpretation for  its values and is dependent on the choice of $\alpha$ in the same way the measure $\Delta \beta(\alpha)$ is.  On the other hand, a wider range of statistical methods may be used for analysis of $\Delta m(\alpha)$ what facilitates the calculations, increases the number of research issues that can be addressed so far and thus increases the applicability of $\Delta m(\alpha)$. Here we may also expect $\Delta m(\alpha)$ to outperform $\Delta \beta(\alpha)$ when quantifying low or high levels of 5hmC, just like $m$-value outperforms $\beta$ in such situations \cite{pan}.

%%%%%%%%%%%%%%%%%%%%%%%%%%%%%%%%%%%%%%%%%%%%%%%%%%%%%%%%%%%%%%%%%%%%
%%%%%%%%%%%%%%%%%%%%%%%%%%%%%%%%%%%%%%%%%%%%%%%%%%%%%%%%%%%%%%%%%%%%

\vspace{0.2cm}
\noindent The most crucial characteristic of the hydroxymethylation  measures $\Delta \beta(\alpha)$ and $\Delta m(\alpha)$ introduced above is their dependence on the choice of the correction term $\alpha$ which also impacts the set of CpGs being selected as those with a substantial level of 5hmC. To address this issue, and eventually to introduce an alternative 5hmC measure without such $\alpha$ dependence, we first analyze the behavior of $\Delta \beta(\alpha)$ and $\Delta m(\alpha)$  as $\alpha$ varies. 
\begin{figure}[htp!]
        \centering
    \centering
  \includegraphics[width=0.9\linewidth, height=0.345\textheight]{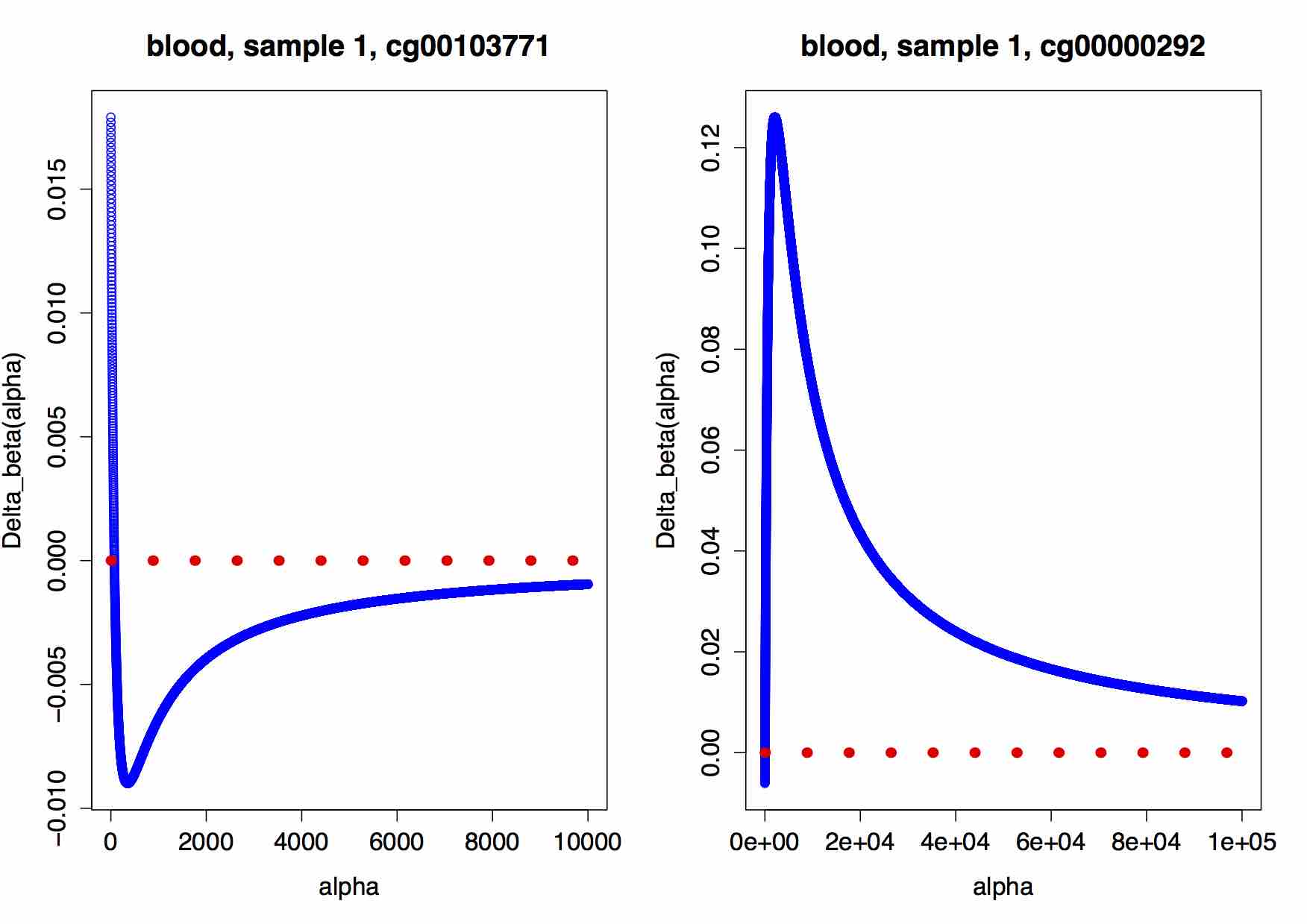}
        \caption{Convergence of $\Delta \beta(\alpha) $  for increasing $\alpha$: a real data example for a given sample, tissue and  CpG.}
        \label{conv0}        
\end{figure}

\vspace{0.25cm}
\begin{figure}[htp]
\hspace{2.3cm} \includegraphics[width=0.82\linewidth, height=0.37\textheight]{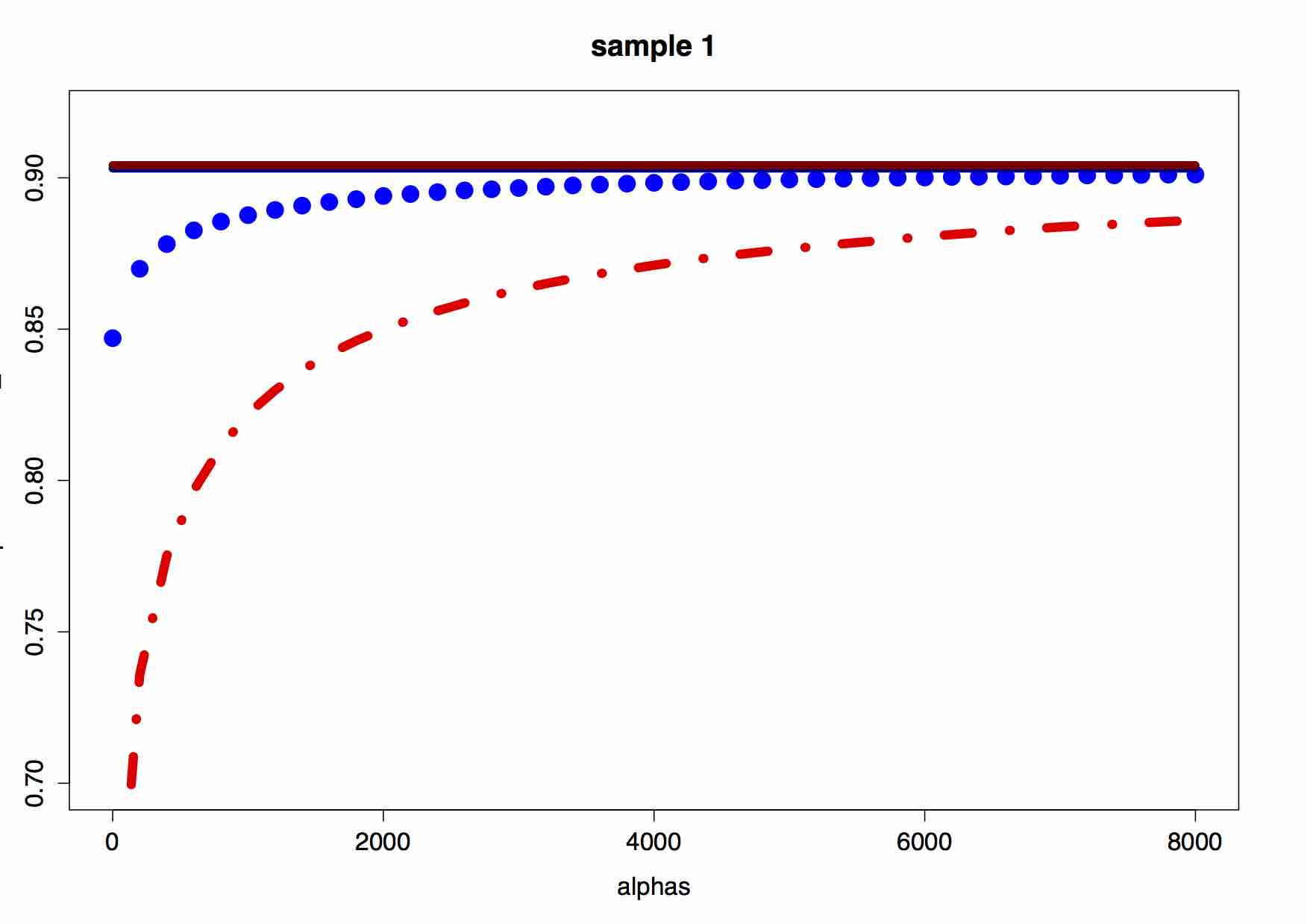}
        \caption{Percentage of CpGs satisfying $\Delta \beta(\alpha) > 0$: a real data example for a given sample (sample 1) and  both tissues.  The blue line corresponds to brain tissue, the red line corresponds to blood tissue; here we consider $\alpha=0, 10, \ldots, 8000$.}
        \label{sub33}
\end{figure}

\vspace{0.2cm}

\noindent As follows from (\ref{b112}), $\Delta \beta(\alpha)$ converges to zero as $\alpha$ increases; see the Appendix \ref{sec2} for a discussion and Figure \ref{conv0} for a graphical presentation of this convergence result. In practice, this result will imply that, for increasing $\alpha$, the range of possible $\Delta \beta(\alpha)$ values will narrow; e.g., the {\it tissue effect} as observed in terms of the corresponding $\Delta \beta(\alpha)$ values may become less observable.

\begin{figure}[htp]
\hspace{1.0cm} \includegraphics[width=0.88\linewidth, height=0.37\textheight]{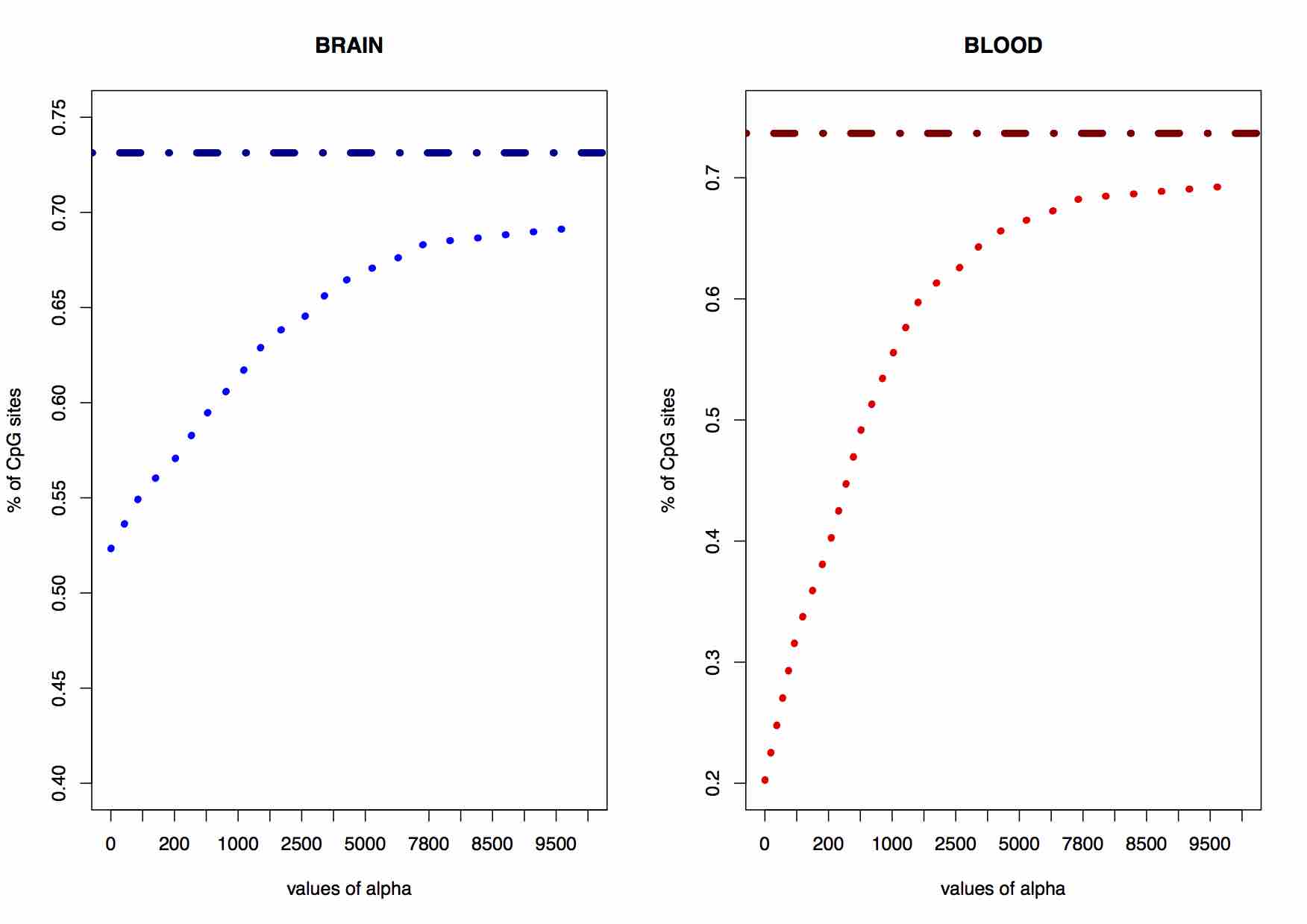}
        \caption{Percentage of CpGs satisfying $\Delta \beta(\alpha) > 0$ over all given samples: a real data example for both tissues and $\alpha=0, 10, \ldots, 10000$.  The blue curve corresponds to brain tissue, the red curve corresponds to blood tissue. The dark blue straight line corresponds to the percentage of CpG sites satisfying $M_{BS}>M_{oxBS}$ over all given samples for brain tissue; the dark red straight line corresponds to the analogous percentage for blood tissue.}
        \label{sub3335}
\end{figure}

\vspace{0.35cm}
\noindent With the convergence result for the measure $\Delta \beta(\alpha)$ obtained above, the crucial question concerning an impact of  increasing values of $\alpha$ on the percentage of CpGs  satisfying the condition $\Delta \beta(\alpha)>0$ for a given sample arises. As already demonstrated by Figure \ref{f055} and now by Figure \ref{sub33}, this percentage converges to a positive constant value as $\alpha$ increases; standard computations verify this limit value to be just the percentage of CpGs  satisfying $M_{BS} > M_{oxBS}$ for a given sample.\footnote{The considered limit value will actually be given by the union of the following two sets $$ \{M_{BS}>M_{oxBS}\}\cup \{M_{BS}=M_{oxBS}, U_{BS}<U_{oxBS}\},$$ but for simplicity we ignore the latter set due to its evident irrelevance for the quantification of 5hmC level and the fact that the percentage of all CpG sites in our real data sets simultaneously satisfying the conditions $$M_{BS}=M_{oxBS}, U_{BS}<U_{oxBS}$$ for any given sample ranges between $0.0002\%$ and $0.1\%$.}  The same convergence result holds while considering the percentage of CpGs which satisfy the condition $\Delta \beta(\alpha)>0$  over all four given samples: for an illustration see Figure \ref{sub3335}.

%\vspace{0.25cm}

\begin{center}
\begin{figure}[htp]
 \hspace{0.55cm}\includegraphics[width=0.93\linewidth, height=0.38\textheight]{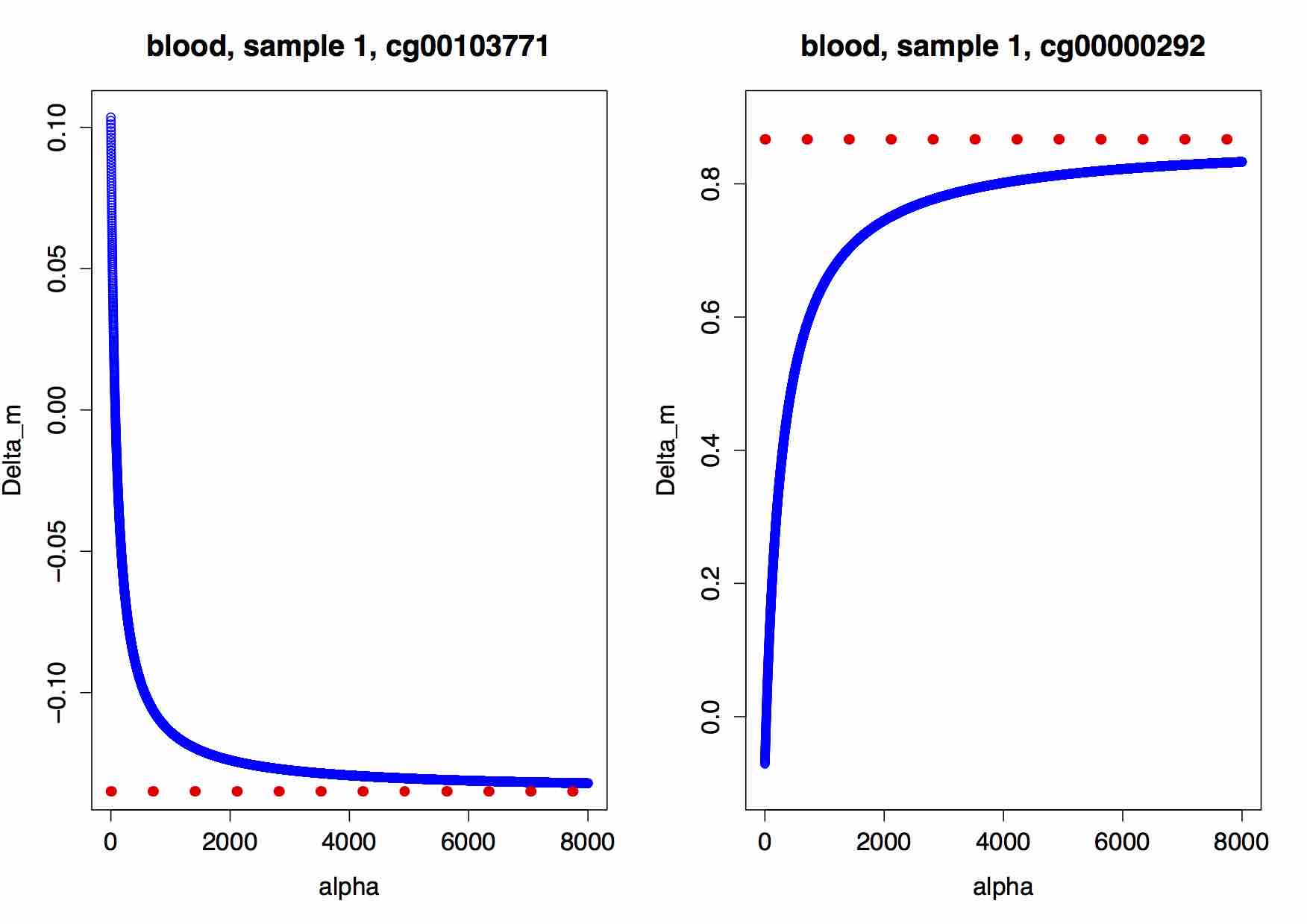}
        \caption{Convergence of $\Delta m(\alpha) $  for increasing $\alpha$: a real data example, for a given sample,  CpG and blood tissue. \vspace{0.5cm}}
        \label{f111bn}
\end{figure}
\end{center}

\vspace{-1.4cm}
 \noindent Also the convergence result obtained for the measure $\Delta \beta(\alpha)$ is transferable to $\Delta m (\alpha).$ In particular, Figure \ref{f111bn} shows that  $\Delta m (\alpha)$  converges to some constant value as $\alpha$ goes to infinity. This result is in accordance with the corresponding convergence result as obtained for  $\Delta \beta(\alpha).$ The only difference is that in case of $\Delta \beta(\alpha)$ this limit will always be zero, independently of the CpGs, sample, and tissue chosen, whereas in case of $\Delta m (\alpha)$ this limit depends on the CpG, sample,  and tissue under consideration.

\begin{figure}[htp]
    \hspace{1.1cm} \includegraphics[width=0.9\linewidth, height=0.35\textheight]{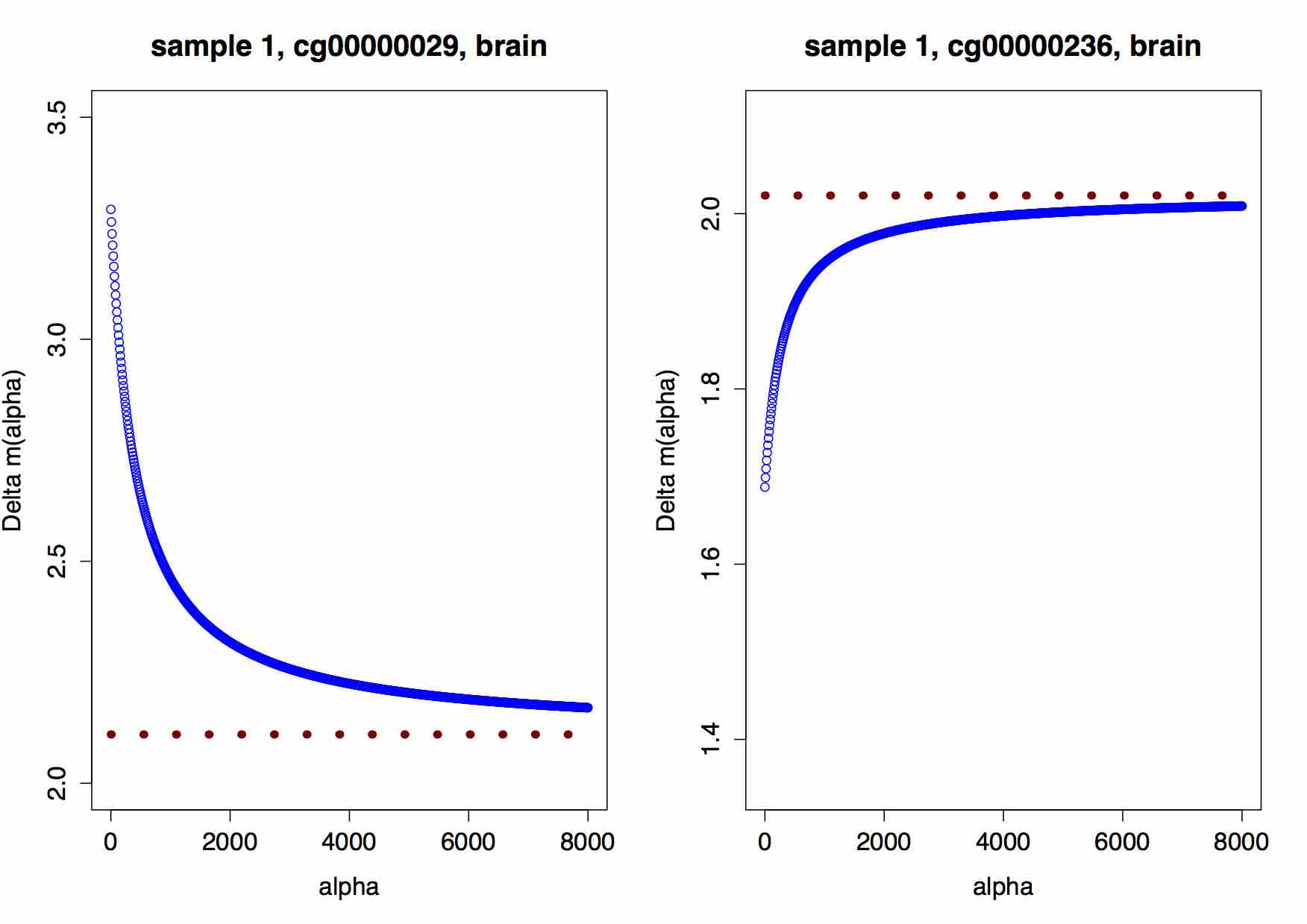}
        \caption{Measures $\Delta m(\alpha) $ (the blue line)  versus $\Delta m^{\infty}$ (the red line) in case when $U_{oxBS} > U_{BS}$  (left-hand panel) and in case when   $U_{oxBS} < U_{BS}$  (right-hand panel). \vspace{0.5cm}}
        \label{vergl}
 \end{figure}

\vspace{0.31cm}
\noindent Inspired by the convergence results  obtained for the measures $\Delta \beta(\alpha)$ and  $\Delta m(\alpha)$ and trying to overcome the dependence of these measures on the choice of the correction term $\alpha$, we propose 
\begin{eqnarray}\label{inf}
\Delta m ^{\infty} = \log_2 \frac{M_{BS}}{M_{oxBS}} (=  \lim_{ \alpha \uparrow  \infty}  \Delta m( \alpha))
\end{eqnarray}
as alternative hydroxymethylation measure. Note that this measure is well-defined for all CpGs satisfying $M_{BS} >0$ and $M_{oxBS}> 0$ simultaneously. 
\vspace{0.3cm}
\\
The main advantage of the measure $ \Delta m ^{ \infty}$ in comparison with the measure $\Delta m(\alpha)$ is its  complete independence of the correction term $ \alpha$; this fact makes the performance of $ \Delta m ^{ \infty}$  more robust. Furthermore, the outcomes of this measure have a very intuitive interpretation. Indeed, we get $\Delta m^{\infty} > 0$ if  $M_{BS} > M_{oxBS}$, i.e., if the global methylated intensity $M_{BS}$ exceeds the "adjusted" methylated intensity $M_{oxBS}$. In all other cases, we will have  $\Delta m ^{\infty} \leq 0;$  for instance, $ \Delta m ^{ \infty} = 0$ implies $M_{BS}=M_{oxBS}$, which  can be interpreted as   \lq no 5hmC observed\rq.
\vspace{0.3cm}
\\
Next, let us address the  possible  relation between the measures  $ \Delta m ^{ \infty}$ and $\Delta m(\alpha)$.  In the context of a single CpG,  simple computations, which are based on the convergence result  for $\Delta m(\alpha)$  obtained above, show that for all CpGs with $$U_{oxBS} \geq U_{BS}$$   the value of $\Delta m(\alpha)$ will be at least as large as the value of $\Delta m ^{\infty}$, for any $\alpha >0$;  in all other case  we will get $\Delta m (\alpha) < \Delta m ^{\infty}$; for an illustration see Figure \ref{vergl}.   \vspace{0.3cm}
\\
 
 \vspace{-0.5cm}
\noindent The final question that is most crucial in the context of the selection procedure concerns a relation between the subsets of CpGs satisfying  $\Delta m(\alpha)>0$ and $\Delta m^{\infty}>0$ respectively. To address this question, we first recall that, for a given sample,  our previous discussion (see Figure \ref{sub33}) has shown that the subset of CpGs detected by the measure $\Delta m^{\infty}$ as those with a substantial level of 5hmC represents the "limiting" subset  for a sequence of subsets of CpGs selected by the measure $\Delta m(\alpha)$ for increasing  $\alpha$. To formalize this result, for a given sample  we divided the set of all CpGs in several disjoint subsets, and showed that, for  increasing $\alpha$, the union of these subsets converges to the subset of CpGs satisfying $M_{BS}>M_{oxBS};$ see the Appendix \ref{infty}. 
 \vspace{0.3cm}

\noindent To summarize the discussion of the present section, we state that,   compared to the measures $\Delta \beta(\alpha)$ and $\Delta m(\alpha)$, the measure $\Delta m^{\infty}$ can have an advantage for quantifying 5hmC levels, in particular,  due to its intuitive interpretation and independence of the choice of the correction term $\alpha$. On the other hand, this measure does not take into account the unmethylated intensities $U_{BS}$ and $U_{oxBS}$. This  may become an issue even if the role of these intensities in detection/quantification of the 5hmC levels has not been clarified yet. Finally, the measure $\Delta m^{\infty}$ will detect a greater number of CpGs as exhibiting a positive level of 5hmC and thus being relevant for the further analysis than both other 5hmC measures will. In this sense, the measure $\Delta m^{\infty}$ is probably more conservative compared to the measures $\Delta m(\alpha)$ and $\Delta \beta(\alpha)$. 
 
 %%%%%%%%%%%%%%%%%%%%%%%%%%%%%%%%%%%%%%%%%%%%%%%%%%%%%%%%%%%%%%%%%%%%%%%%%%
 %%%%%%%%%%%%%%%%%%%%%%%%%%%%%%%%%%%%%%%%%%%%%%%%%%%%%%%%%%%%%%%%%%%%%%%%%%

 \vspace{0.3cm}
\noindent As already mentioned above, the measure  $\Delta m^{\infty}$ does not take into account unmethylated intensities $U_{BS}$ and $U_{oxBS}$. Even if the potential consequences of such modeling are unclear yet, we are going to address this issue by  proposing  another measure for  the quantification of the 5hmC level which  takes also  the unmethylated intensities in account. It is defined as\begin{eqnarray}\label{dh}
\Delta h = 1- \frac{M_{oxBS}+U_{oxBS}}{M_{BS}+U_{BS}},
\end{eqnarray}
where  $M_{BS}+U_{BS}$ is the global methylation level obtained from the {\it BS-seq} method and $M_{oxBS}+U_{oxBS}$ is the global methylation level derived by means of the {\it oxBS-seq} method. Here we have to assume that $M_{BS}+U_{BS}> 0$; all CpGs with $M_{BS}+U_{BS}=0$ have to be exclude from the analysis as exhibiting a measurement error. Note that the measure $\Delta h$ can be obtained directly from the measured data, since both quantities $M_{BS}+U_{BS}$ and $M_{oxBS}+U_{oxBS}$ are immediately observed. Thus, no further data transformations will be necessary, which reduces the possibility of computational errors. 
\vspace{0.3cm}
\\
As follows from its definition, for CpGs with a positive level of 5hmC the measure $\Delta h$ must range between 0 and 1. The CpGs with $\Delta h <0$ are  to be considered as containing only a unsubstantial level of 5hmC;  the CpGs with $\Delta h >1$ are  to be seen as a result of measurement noise.
\vspace{0.3cm}
\\
When interpreting the values of $\Delta h$ in the context of the observed 5hmC level, we can state the following: Intuitively,   for a given sample and CpG, the condition $\Delta h \approx 0$ implies $M_{BS}+U_{BS} \approx M_{oxBS}+U_{oxBS}$ and thus the global 5hmC level is negligible in such situations. In cases, where  $\Delta h \approx 1$ we get  $M_{oxBS} +U_{oxBS}\ll  M_{BS}+U_{BS}$ and thus  the global 5hmC level  has to be high. In particular, larger values of $\Delta h$ correspond to larger percentages of the global 5hmC levels. Altogether, we can interpret $\Delta h$ as {\it the proportion/percentage of 5hmC in the global methylation}.
\vspace{0.3cm}
\\
Next, let us analyze whether positive values of the measure $\Delta h$ lead to positivity of other 5hmC measures introduced above, and vice versa; this relation is particularly important in the selection procedure. As (\ref{dh}) implies, the inequality $\Delta h >0$ holds if $$M_{BS}+U_{BS} > M_{oxBS}+U_{oxBS}.$$ However, the latter inequality is not sufficient to make a statement about the sign of the measures $\Delta\beta(\alpha)$, $\Delta m(\alpha)$, and $\Delta m^{\infty},$ and additional assumptions are needed. For more details see Appendix \ref{wd100}. 
\vspace{0.3cm}
\\
Finally, let us  impose a lower threshold for the measure $\Delta h$ and consider possible consequences. For instance, we can assume that a given CpG exhibits a substantial level of 5hmC only if the corresponding value of $\Delta h$ satisfies the inequality $\Delta h > c$  for some constant  $c>0$. Then a simple calculation shows that the global 5hmC level $M_{5hmC}+U_{5hmC}$ will be bounded from below by the quantity $c\dot(M_{BS}+U_{BS})$. In other words, {\it by imposing a threshold on the measure $\Delta h$ we also set a lower bound for the global 5hmC level.}
\vspace{0.3cm}
\\ 
Altogether, in this section we show that the application of the measure $\Delta h$ for the quantification of 5hmC levels can be of advantage, since this measure overcomes the limitation of the 5hmC measures proposed before, and, in particular, does not depend on the choice of a correction term $\alpha$. It has an intuitive interpretation of the outcomes in terms of the observed 5hmC level, and can be easily computed from the measured data.

  %%%%%%%%%%%%%%%%%%%%%%%%%%%%%%%%%%%%%%%%%%%%%%%%%%%%%%%%%%%%%%%%%%%%%%%%%%%%%%%%%
  %%%%%%%%%%%%%%%%%%%%%%%%%%%%%%%%%%%%%%%%%%%%%%%%%%%%%%%%%%%%%%%%%%%%%%%%%%%%%%%%%

  \section{Similarity analyses}

 In order to compare the outputs of the proposed 5hmC measures without making any statement about their optimality/performance {\it similarity analyses} can be used; the main tool of such similarity analyses is  a {\it similarity measure}. The aim of the present section is to introduce a similarity measure which can be used  for pairwise comparison of the considered 5hmC measures, then apply this similarity measure to our real data sets and discuss the observed results.
 \vspace{0.25cm}
 \\
As a reminder: we have two real data sets which correspond to brain and whole blood tissue respectively; such a tissues' choice seems to be particularly interesting, since the brain is known to have the highest level of 5hmC, while blood is known to have the lowest, see, e.g.,  \cite{laird}. Each data set  consists of four independent samples; there are four intensity vectors, $$M_{BS}, M_{oxBS}, U_{BS}, U_{oxBS}, $$ available for each sample. The data  used for analysis was not normalized.
\vspace{0.25cm}
\\
%%%%%%%%%%%%%%%%%%%%%%%%%%%%%%%%%%%%%%%%%%%%%%%%%%%%%%%%%%%%%%%%%%%%%%%%%%%%%%%%%
%%%%%%%%%%%%%%%%%%%%%%%%%%%%%%%%%%%%%%%%%%%%%%%%%%%%%%%%%%%%%%%%%%%%%%%%%%%%%%%%%
\noindent In order to quantify the similarity of two given 5hmC measures in the context of the screening step, we introduce the { \it similarity measure}  $\mathbb S$ that quantifies the {\it pairwise agreement}  or {\it pairwise similarity} of the proposed 5hmC measures $\Delta \beta(\alpha)$, $\Delta m^{\infty}$ and $\Delta h$.
In particular, for a given CpG\footnote{or for a given sample, if we perform our analysis sample-wise.}  we define
\begin{eqnarray}\label{sm1}
\mathbb S(x_1, x_2)= \frac1n\bigg( \sum_{i=1}^n I_{\{x_1^i>0\}} I_{\{x_2^i>0\}}+ \sum_{i=1}^n I_{\{x_1^i\leq0\}} I_{\{x_2^i\leq0\}}\bigg).
\end{eqnarray}

\noindent Here $x_1, x_2$ denote any two considered  5hmC measures, $n$ is the number of samples under consideration\footnote{or the number of CpGs for a given sample, if we perform our analysis sample-wise.} and $I_{ \{x\}}$ is the indicator function, defined as 
\begin{eqnarray}
I_{\{x>0\}} = \begin{cases} 1 & x >0 \\ 0 & \text{otherwise.} \end{cases}
\end{eqnarray} 
Clearly, the similarity measure $\mathbb S$ in \eqref{sm1} ranges between $0$ and $1$, where $1$ denotes {\it complete similarity} and $0$ denotes {\it complete dissimilarity}.

\begin{figure}[htp!]
    \centering
    \vspace{0.55cm}
\begin{overpic}[width=14cm]{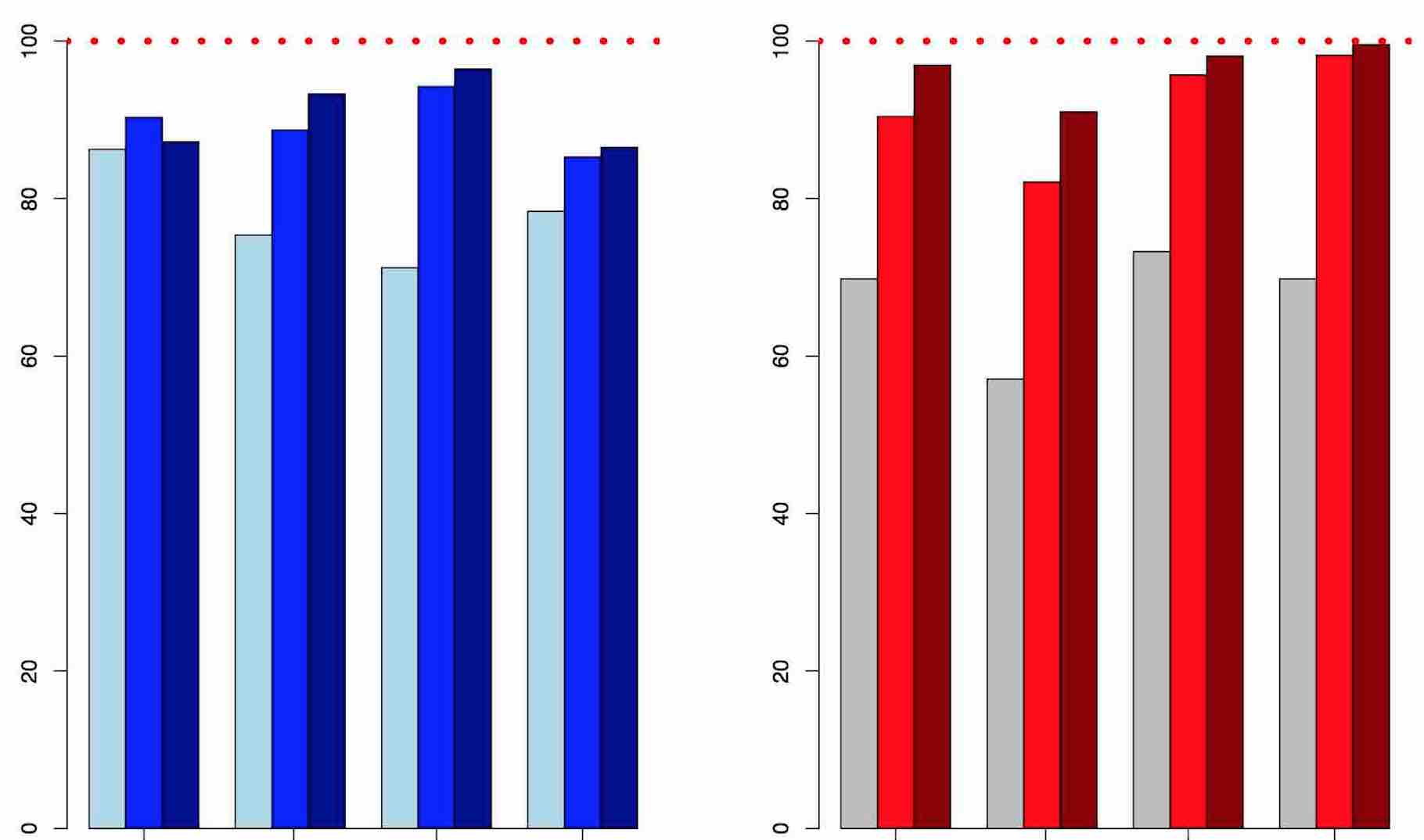}
\put(35,-14){s1}
\put(75,-14){s2}
\put(115,-14){s3}
\put(157,-14){s4}
\put(245,-14){s1}
\put(287,-14){s2}
\put(330,-14){s3}
\put(369,-14){s4}
\put(21,235){brain tissue}
\put(231,235){blood tissue}
\put(75,-35){samples}
\put(290,-35){samples}
 \end{overpic}
\vspace{1.2cm}
\caption{Sample-wise application of the 5hmC measures $\Delta \beta(100), \Delta m^{\infty}$ and $\Delta h$. Both panels present the percentage of CpGs being detected by each of the considered 5hmC measures as exhibiting a positive 5hmC level.   The left-hand panel describes the results for  brain tissue. On this panel, the light blue bars correspond to the $\Delta \beta(100)$ measure, the blue bars to the $\Delta m^{\infty}$ measure and the dark blue bars to the $\Delta h$ measure.   The right-hand panel describes the results for  blood tissue. On this panel, the grey bars correspond to the $\Delta \beta(100)$ measure, the red bars to the $\Delta m^{\infty}$ measure and the dark red bars to the $\Delta h$ measure.\vspace{0.3cm}}
        \label{fig12}

 %   \end{minipage}
\end{figure}

\vspace{0.35cm}
\noindent At the beginning of real data analysis,  we first  applied all three 5hmC measures, $\Delta \beta(100),$ $\Delta m^{\infty}$ and $\Delta h$, on both real data sets and computed the percentage of CpG sites being detected as hydroxymethylated by each of these measures; the results of these computations are presented in Figure \ref{fig12}.   This figure depicts the measure $\Delta h$ as being the most conservative while detecting the hydroxymethylated CpGs, since the percentage of CpGs marked by this measure as being hydroxymethylated and thus  relevant for the further analysis is the largest one.

 \begin{figure}[t!]
    \centering
    \vspace{0.55cm}
\begin{overpic}[width=14cm]{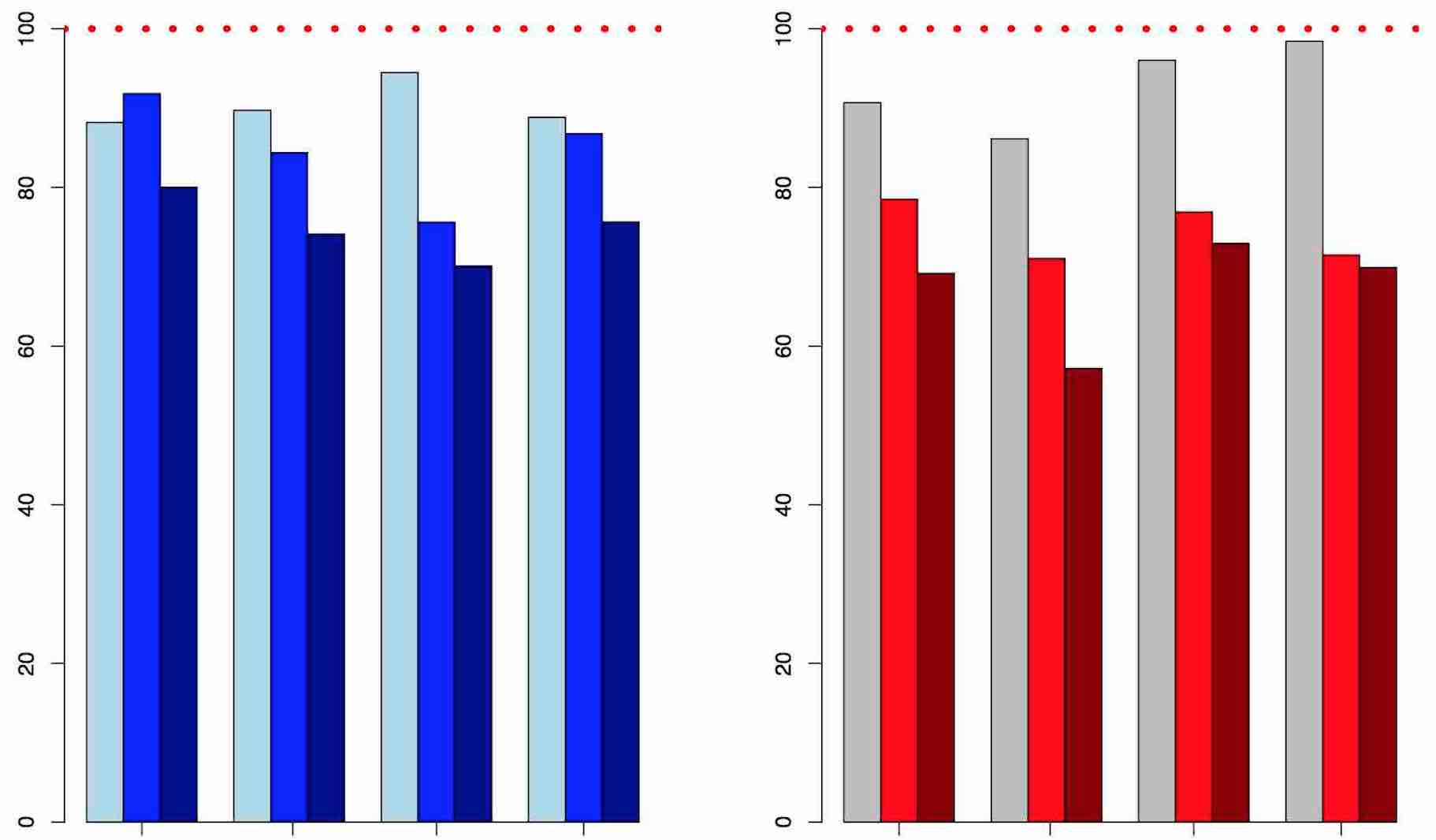}
\put(35,-14){s1}
\put(75,-14){s2}
\put(115,-14){s3}
\put(157,-14){s4}
\put(245,-14){s1}
\put(287,-14){s2}
\put(330,-14){s3}
\put(369,-14){s4}
\put(21,235){brain tissue}
\put(231,235){blood tissue}
\put(75,-35){samples}
\put(290,-35){samples}
 \end{overpic}
\vspace{1.2cm}
       \caption{Application of the similarity measure $\mathbb S$ for the pairwise comparison of the 5hmC measures $\Delta \beta(100), \Delta m^{\infty}$ and $\Delta h$. The similarity was analyzed sample-wise, for each of four given samples. Both panels present the percentage of CpGs  detected by a pair of the considered 5hmC measures as exhibiting a positive 5hmC level.   The left-hand panel describes the results for  brain tissue. On this panel, the light blue bars correspond to the values of $\mathbb S(\Delta m^{\infty}, \Delta h)$, the blue bars to the  values of $\mathbb S(\Delta m^{\infty}, \Delta \beta(100))$  and the dark blue bars to the values of $\mathbb S(\Delta \beta(100), \Delta h)$.   The right-hand panel describes the results for  blood tissue. On this panel, the grey bars correspond  to the values of $\mathbb S(\Delta m^{\infty}, \Delta h)$, the red bars  to the values of $\mathbb S(\Delta m^{\infty}, \Delta \beta(100))$ and the dark red bars  to the values of $\mathbb S(\Delta \beta(100), \Delta h)$.\vspace{0.15cm}}
        \label{nmb}
        
 %   \end{minipage}
\end{figure}
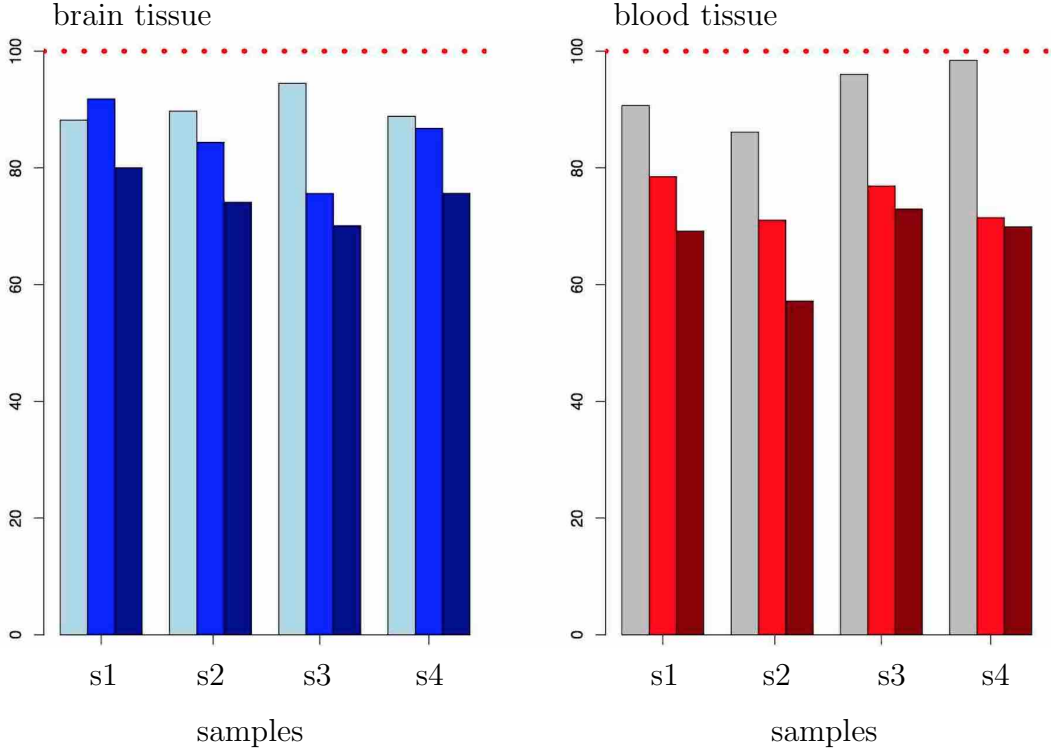

\begin{figure}[t!]
    \centering
    \vspace{-1.2cm}
  \begin{overpic}[width=10cm]{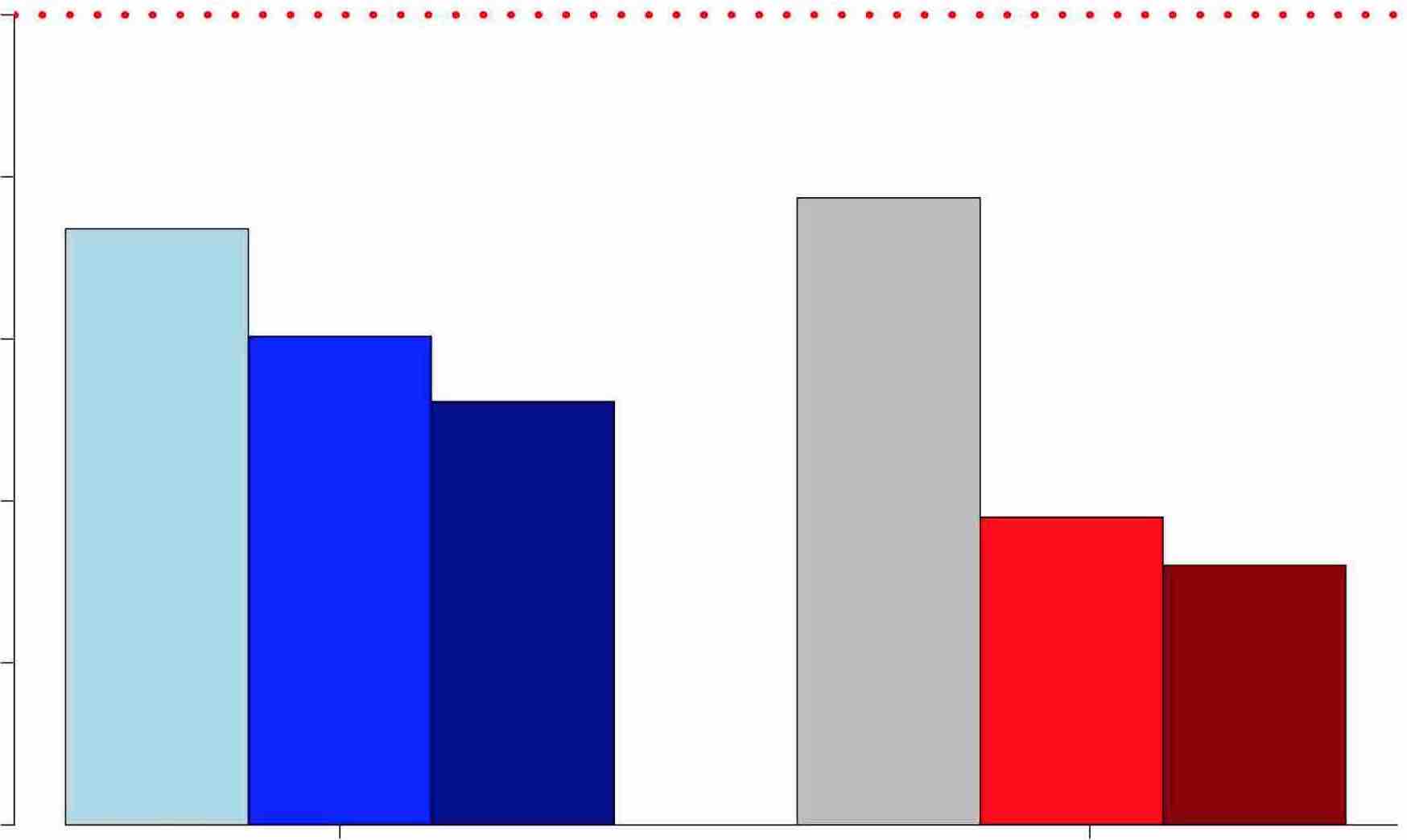}
\put(60,-35){brain tissue}
\put(195,-35){blood tissue}
\put(-15,1){0}
\put(-20,31){20}
\put(-20,65){40}
\put(-20,95){60}
\put(-20,130){80}
\put(-25,160){100}
 \end{overpic}
\vspace{1.2cm}
      \caption{Pairwise complete similarity performed for each given CpG site.  The light blue bar corresponds to the percentage of CpGs with $\mathbb S(\Delta m^{\infty}, \Delta h)=1$ over all given samples, in brain tissue. The blue bar corresponds to the percentage of CpGs with $\mathbb S(\Delta m^{\infty}, \Delta \beta(100))=1$ over all given samples, in brain tissue. The dark blue bar corresponds to the percentage of CpGs with $\mathbb S(\Delta \beta(100), \Delta h)=1$ over all given samples, in brain tissue. The grey bar corresponds to the percentage of CpGs with $\mathbb S(\Delta m^{\infty}, \Delta h)=1$ over all given samples, in  blood tissue. The red bar corresponds to the percentage of CpGs with $\mathbb S(\Delta m^{\infty}, \Delta \beta(100))=1$ over all given samples, in  blood tissue. The dark red bar corresponds to the percentage of CpGs with $\mathbb S(\Delta \beta(100), \Delta h)=1$ over all given samples, in  blood tissue. }
        \label{smm1}        
 %   \end{minipage}
\end{figure}

\begin{figure}[htp!]
    \centering
 \vspace{-0.2cm}
\begin{overpic}[width=10cm]{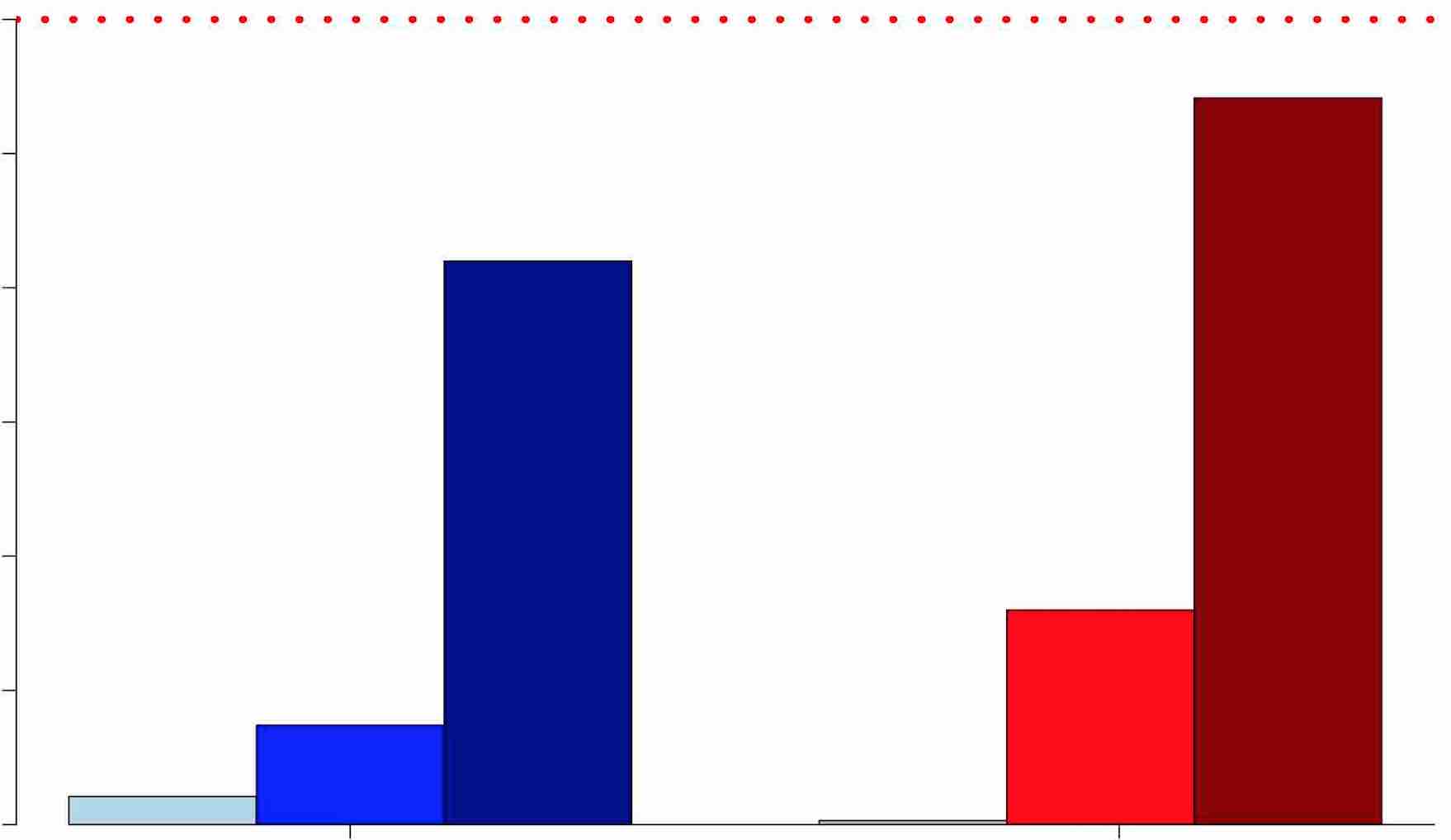}
\put(60,-35){brain tissue}
\put(195,-35){blood tissue}
\put(-15,1){0}
\put(-15,24){1}
\put(-15,52){2}
\put(-15,78){3}
\put(-15,105){4}
\put(-15,130){5}
\put(-15,156){6}
 \end{overpic}
\vspace{1.2cm}
    \caption{Pairwise complete dissimilarity performed for each given CpG site.  The light blue bar corresponds to the percentage of CpGs with $\mathbb S(\Delta m^{\infty}, \Delta h)=0$ over all given samples, in brain tissue. The blue bar corresponds to the percentage of CpGs with $\mathbb S(\Delta m^{\infty}, \Delta \beta(100))=0$ over all given samples, in brain tissue. The dark blue bar corresponds to the percentage of CpGs with $\mathbb S(\Delta \beta(100), \Delta h)=0$ over all given samples, in brain tissue. The grey bar corresponds to the percentage of CpGs with $\mathbb S(\Delta m^{\infty}, \Delta h)=0$ over all given samples, in blood tissue. The red bar corresponds to the percentage of CpGs with $\mathbb S(\Delta m^{\infty}, \Delta \beta(100))=0$ over all given samples, in blood tissue. The dark red bar corresponds to the percentage of CpGs with $\mathbb S(\Delta \beta(100), \Delta h)=0$ over all given samples, in blood tissue.}
        \label{ssmm2}   
\end{figure}

\begin{figure}[htp!]
    \centering
 \vspace{-0.2cm}
\begin{overpic}[width=10cm]{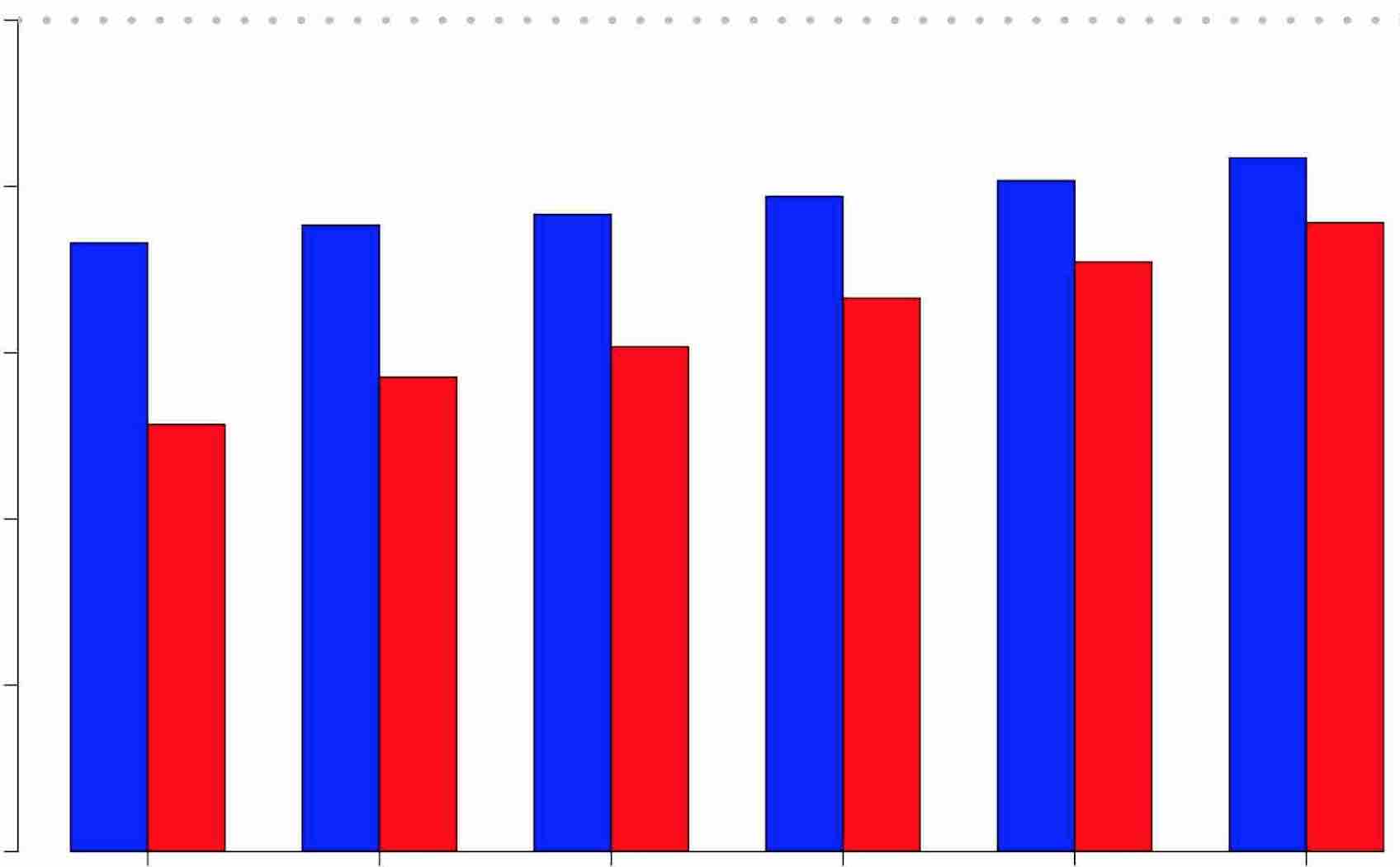}
\put(-15,1){0}
\put(-20,31){20}
\put(-20,65){40}
\put(-20,98){60}
\put(-20,135){80}
\put(-25,170){100}

\put(30,-15){0}
\put(68,-15){100}
\put(115,-15){200}
\put(165,-15){500}
\put(210,-15){1000}
\put(255,-15){2500}
\put(115,-35){values of $\alpha$}

 \end{overpic}
\vspace{1.2cm}
    \caption{Percentage of CpG sites satisfying $\Delta \beta(\alpha)>0$ for increasing $\alpha$ and a given sample (sample 2). The blue bars correspond to brain tissue, the red bars to blood tissue. }
        \label{dep_a}
\end{figure}

\begin{figure}[htp!]
    \centering
 \vspace{-0.2cm}
\begin{overpic}[width=10cm]{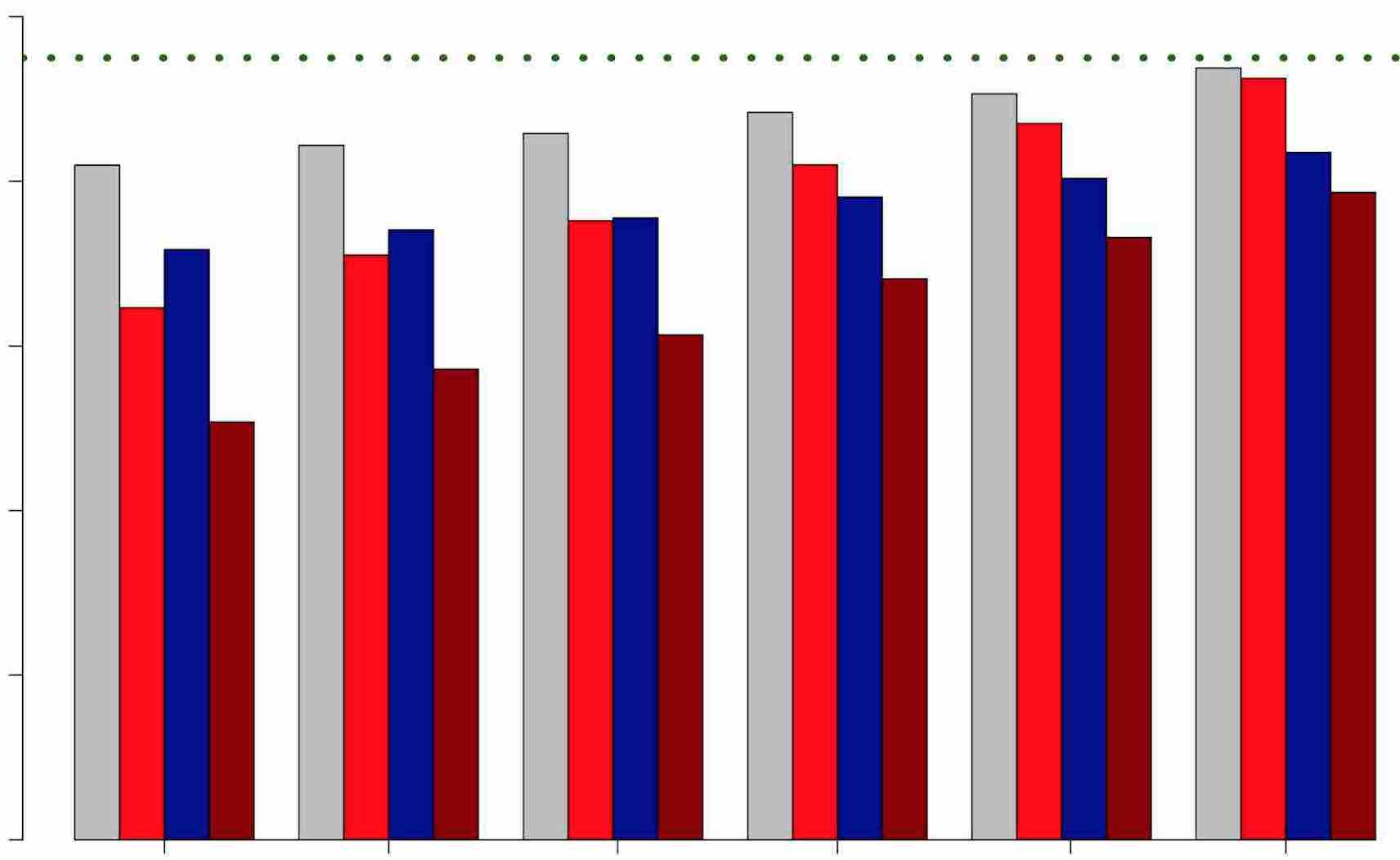}
\put(-15,1){0}
\put(-20,31){20}
\put(-20,65){40}
\put(-20,101){60}
\put(-20,135){80}
\put(-25,164){100}

\put(30,-15){0}
\put(68,-15){100}
\put(115,-15){200}
\put(165,-15){500}
\put(210,-15){1000}
\put(255,-15){2500}
\put(115,-35){values of $\alpha$}
\end{overpic}
\vspace{1.2cm}
     \caption{Application of the similarity measure $\mathbb S$ for the pairwise comparison of the 5hmC measures $\Delta \beta(\alpha), \Delta m^{\infty}$ and $\Delta h$; the computation were performed for increasing values of $\alpha$. The similarity was analyzed over all given CpGs of a given sample (sample 2). Single bars present the percentage of CpGs being detected by a pair of the considered 5hmC measures as exhibiting a positive 5hmC level.   The grey bars correspond to the comparison of the 5hmC measures $\Delta \beta(\alpha)$ and $\Delta m^{\infty}$ on brain tissue; the red bars describe the results of the analogous comparison on blood tissue. The dark blue bars correspond to the comparison of the 5hmC measures $\Delta \beta(\alpha)$ and $\Delta h$ on brain tissue; the dark red bars describe the results of the analogous comparison on blood tissue. }
        \label{dep_a_2}  
\end{figure}
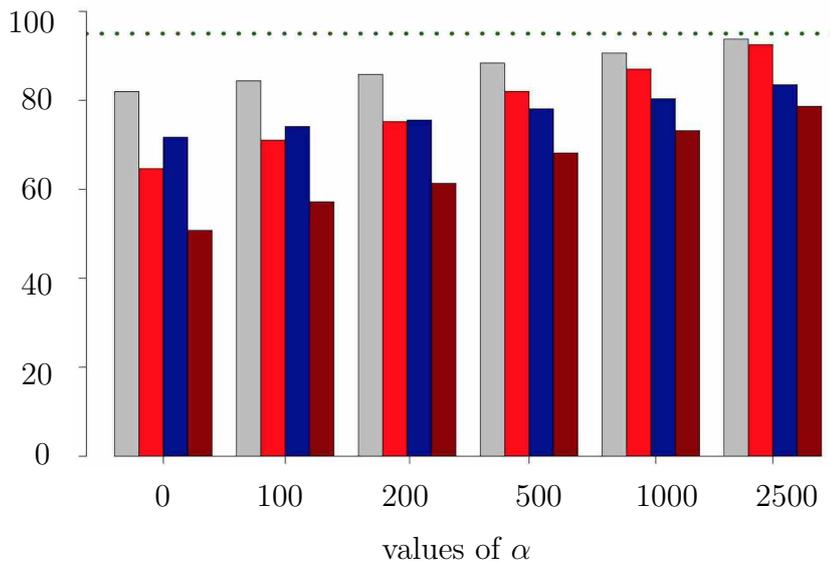

\vspace{0.1cm}
\noindent Next, we applied the measure $\mathbb S$  in order to address the pairwise similarity of the 5hmC measures $\Delta \beta(100), \Delta m^{\infty}, \Delta h$  for each given sample. The obtained results are presented in Figure \ref{nmb}.  In that figure, the highest similarity can be observed between the 5hmC measures $\Delta m^{\infty}$ and $\Delta h$; the measures $\Delta \beta(100)$ and $\Delta h$ demonstrate the weakest similarity in terms of the similarity measure $\mathbb S.$
\vspace{0.35cm}
\\
In order to discuss the 5hmC measures with respect to their pairwise complete similarity as well as complete dissimilarity, we performed a pairwise comparison of these measures for each given CpG. The results of such a comparison are presented in Figures \ref{smm1} and \ref{ssmm2}. 
%\vspace{0.001cm}
\\
In particular, Figure   \ref{smm1} demonstrates the highest level of complete similarity to be observed between the 5hmC measures $\Delta m^{\infty}$ and $\Delta h$, independent of the considered tissue.  On the other hand, Figure   \ref{ssmm2} shows the highest level of complete dissimilarity between the 5hmC measures $\Delta \beta(100)$ and $\Delta h$, independent of the considered tissue; the 5hmC measures $\Delta m^{\infty}$ and $\Delta h$ demonstrate the lowest level of complete dissimilarity in that  figure. In total, $\Delta m^{\infty}$ and $\Delta h$ seem to be the two most similar 5hmC measures  in the context of the similarity measure $\mathbb S.$
\vspace{0.35cm}
\\
\noindent As discussed in the previous section, one of the main limitations  of the measure $\Delta \beta(\alpha)$ is its dependence on the correction term $\alpha.$ Thus we address this dependence in the context of the similarity measure $\mathbb S.$ As already mentioned earlier, the number of CpG sites satisfying $\Delta \beta(\alpha)>0$  grows with increasing $\alpha;$ see Figure \ref{dep_a} for an illustration. 
\vspace{0.35cm}
\\
\noindent Changes in the pairwise similarity between two given 5hmC measures, as computed for all CpGs of a given sample, are described in Figure \ref{dep_a_2}.  As observed on that picture, pairwise similarity between the 5hmC measure $\Delta \beta(\alpha)$ and the remaining two 5hmC measures $\Delta m^{\infty}$ and $\Delta h$ increases with increasing $\alpha.$

\vspace{0.3cm}

\begin{figure}[htp!]
    \centering
 \vspace{-0.1cm}
\begin{overpic}[width=10cm]{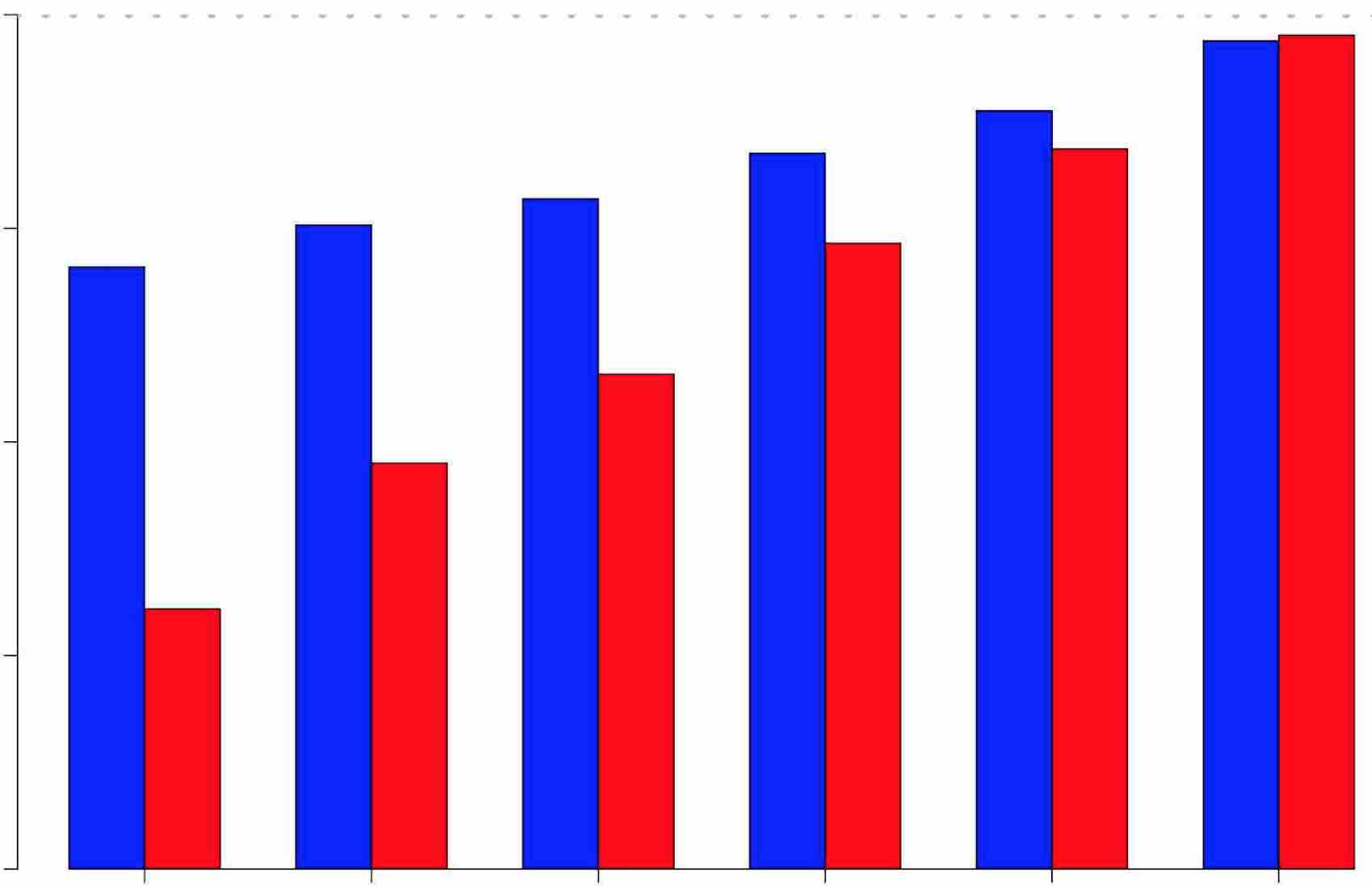}
\put(-15,1){0}
\put(-20,44){20}
\put(-20,89){40}
\put(-20,132){60}
\put(-20,175){80}

\put(30,-15){0}
\put(68,-15){100}
\put(115,-15){200}
\put(165,-15){500}
\put(210,-15){1000}
\put(255,-15){2500}
\put(115,-35){values of $\alpha$}
\end{overpic}
\vspace{1.2cm}
    \caption{Pairwise complete similarity computed for each given CpG site, both given tissues and increasing values of $\alpha.$  The blue bars correspond to the percentage of CpGs with $\mathbb S(\Delta m^{\infty}, \Delta \beta(\alpha))=1$ over all given samples, in brain tissue. The red bars correspond to the percentage of CpGs with $\mathbb S(\Delta m^{\infty}, \Delta \beta(\alpha))=1$ over all given samples, in blood tissue. Note that complete similarity increases with increasing $\alpha.$}
        \label{dep_a_4}
\end{figure}
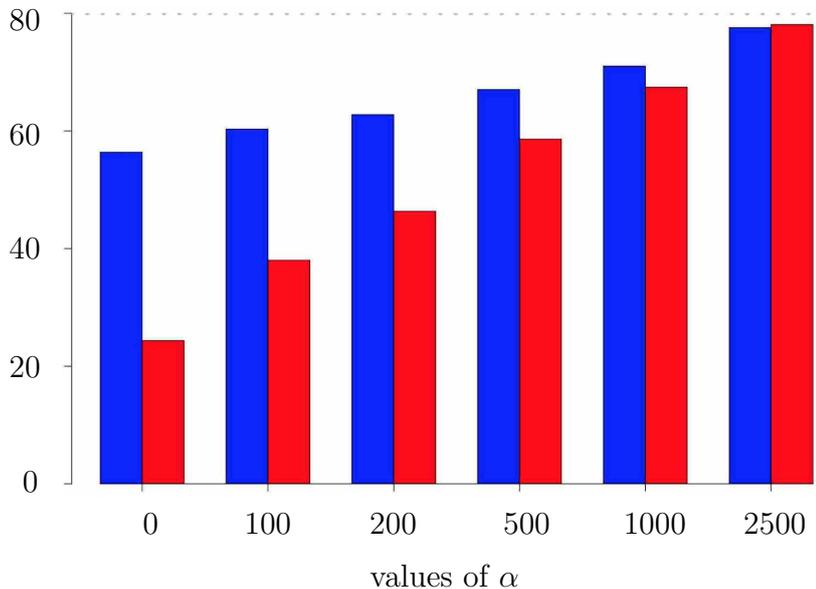

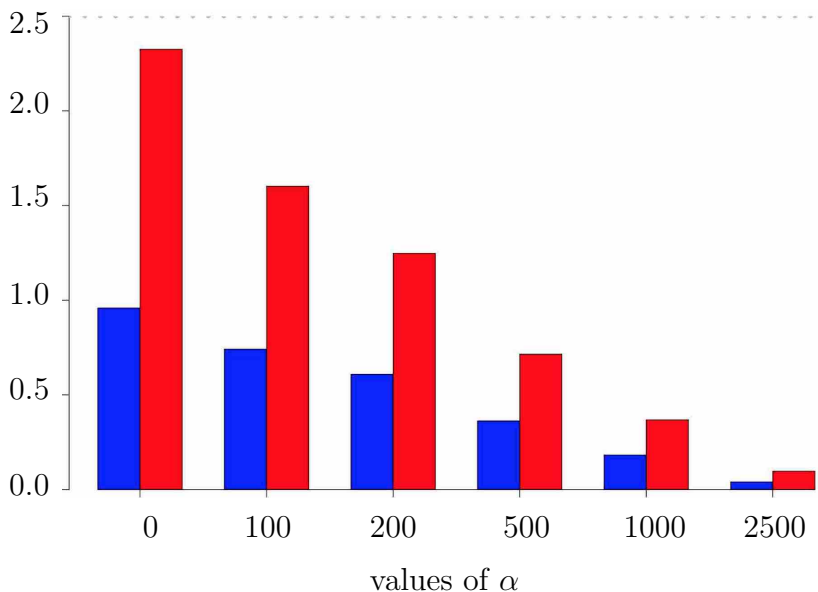
\begin{figure}[htp!]
    \centering
 \vspace{-0.4cm}
\begin{overpic}[width=10cm]{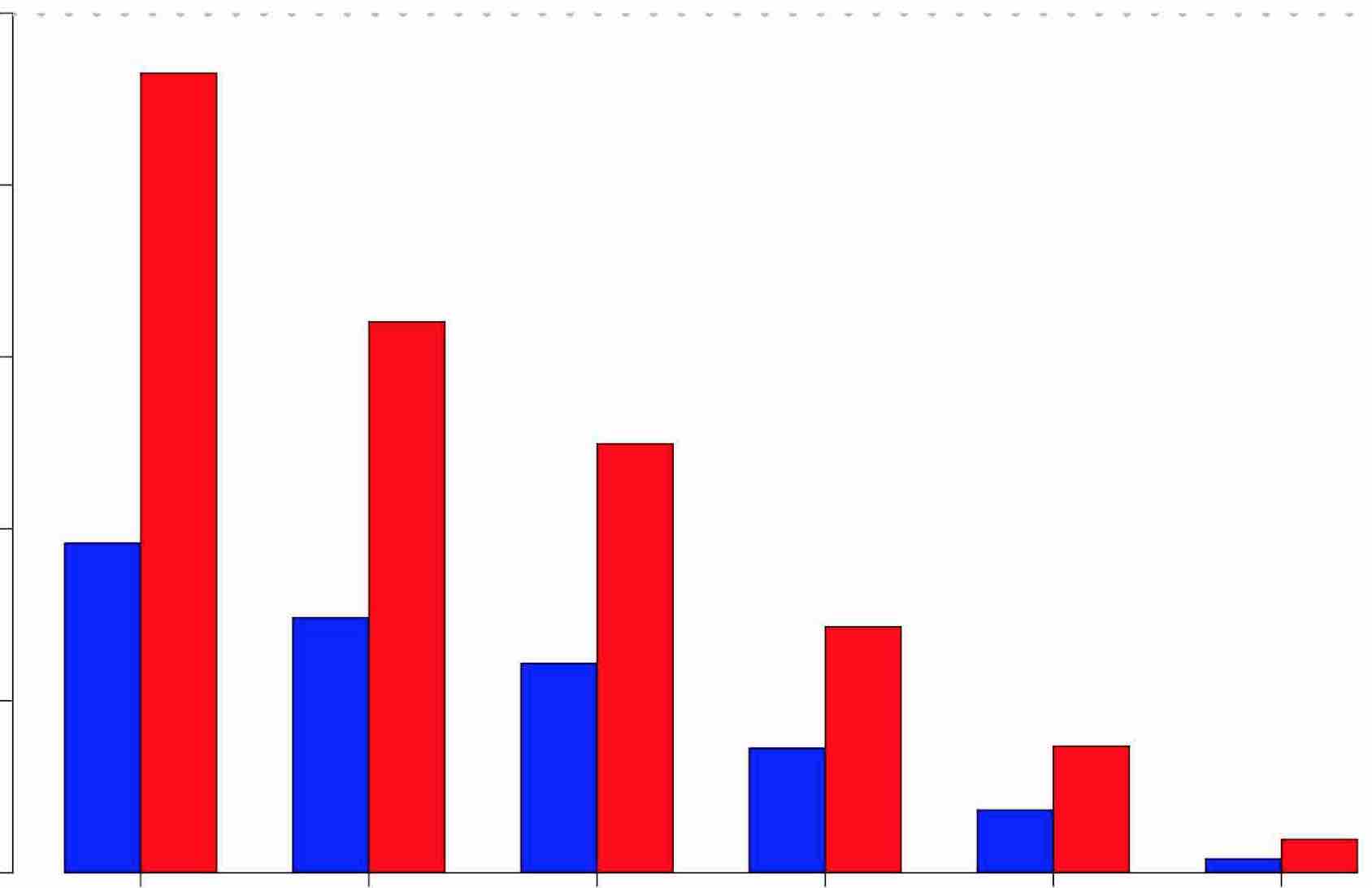}
\put(-20,1){0.0}
\put(-20,37){0.5}
\put(-20,73){1.0}
\put(-20,110){1.5}
\put(-20,145){2.0}
\put(-20,175){2.5}

\put(30,-15){0}
\put(68,-15){100}
\put(115,-15){200}
\put(165,-15){500}
\put(210,-15){1000}
\put(255,-15){2500}
\put(115,-35){values of $\alpha$}
\end{overpic}
\vspace{1.2cm}
    \caption{Pairwise complete dissimilarity computed for each given CpG site, both given tissues and increasing values of $\alpha.$  The blue bars correspond to the percentage of CpGs with $\mathbb S(\Delta m^{\infty}, \Delta \beta(\alpha))=0$ over all given samples, in brain tissue. The red bars correspond to the percentage of CpGs with $\mathbb S(\Delta m^{\infty}, \Delta \beta(\alpha))=0$ over all given samples, in blood tissue. Note that complete dissimilarity decreases with increasing $\alpha.$}
        \label{dep_a_5}
\end{figure}

\clearpage
\noindent Possible impact of the increasing values of $\alpha$ on  the pairwise complete similarity as well as complete dissimilarity between the considered 5hmC measures is illustrated in Figures \ref{dep_a_4} and \ref{dep_a_5}.  In particular, Figure \ref{dep_a_4}  suggests that the pairwise complete similarity increases with increasing $\alpha$; Figure \ref{dep_a_5}  demonstrates decreasing pairwise dissimilarity as $\alpha$ grows.

%%%%%%%%%%%%%%%%%%%%%%%%%%%%%%%%%%%%%%%%%%%%%%%%%%%%%%%%%%%%%%%%%%%%%%%%%%%%%%%%%
%%%%%%%%%%%%%%%%%%%%%%%%%%%%%%%%%%%%%%%%%%%%%%%%%%%%%%%%%%%%%%%%%%%%%%%%%%%%%%%%%

\clearpage

%%%%%%%%%%%%%%%%%%%%%%%%%%%%%%%%%%%%%%%%%%%%%%%%%%%%%%%%%%%%%%%%%%%%%%%%%%
%%%%%%%%%%%%%%%%%%%%%%%%%%%%%%%%%%%%%%%%%%%%%%%%%%%%%%%%%%%%%%%%%%%%%%%%%%

%%%%%%%%%%%%%%%%%%%%%%%%%%%%%%%%%%%%%%%%%%%%%%%%%%%%%%%%%%%%%%%%%%%%%%%%%%%%%%%%%
%%%%%%%%%%%%%%%%%%%%%%%%%%%%%%%%%%%%%%%%%%%%%%%%%%%%%%%%%%%%%%%%%%%%%%%%%%%%%%%%%

%%%%%%%%%%%%%%%%%%%%%%%%%%%%%%%%%%%%%%%%%%%%%%%%%%%%%%%%%%%%%%%%%%%%%%%%%%%%%%%%%
%%%%%%%%%%%%%%%%%%%%%%%%%%%%%%%%%%%%%%%%%%%%%%%%%%%%%%%%%%%%%%%%%%%%%%%%%%%%%%%%%

%%%%%%%%%%%%%%%%%%%%%%%%%%%%%%%%%%%%%%%%%%%%%%%%%%%%%%%%%%%%%%%%%%%%%%%%%%%%%%%%%
%%%%%%%%%%%%%%%%%%%%%%%%%%%%%%%%%%%%%%%%%%%%%%%%%%%%%%%%%%%%%%%%%%%%%%%%%%%%%%%%%

\vspace{1cm}
\section{Conclusion}
Presently, the most applied measure for quantifying a level of 5hmC is the  measure $\Delta \beta$ introduced in \cite{stewart, field}. In particular, this measure is applied  for detection of the CpG sites with a positive level of 5hmC; those CpG sites are then considered to be relevant for the further (also statistical) analysis. 
\vspace{0.3cm}
\\
In this paper we first perform a detailed analysis of  $\Delta \beta$, both analytically and data-based, and discuss a number of limitations which make the application of this measure while selecting hydroxymethylated CpG sites debatable.  To overcome these limitations, we then propose two alternative 5hmC measures which can be used instead of $\Delta \beta.$  The properties for these 5hmC measures as well as their relation to the initial measure $\Delta \beta$ are also discussed, both analytically and on real data. We also propose {\it similarity analyses} which can be used in order to compare the considered 5hmC measures in a certain sense.
\vspace{0.03cm}
\\
Note that our real data analyses are based on the raw data, i.e. the data without any normalization or pre-processing procedures before the down-stream statistical analysis. This differs from the common procedures proposed in  \cite{stewart, field} where only normalized data was applied. To justify our method, we first recall that the most of our results were initially obtained analytically and only then confirmed by means of the data analysis.  Thus, similar results can be expected to hold in case of the normalized data. On the other hand, the normalization of the data may give raise to a number of questions such as a possible impact of normalization while quantifying the 5hmC level. This impact has not been investigated so far, and  it is not clear whether certain results derived on the normalized  data will be just due to the normalization itself. The choice of an appropriate normalization method remains an open issue as well.

%%%%%%%%%%%%%%%%%%%%%%%%%%%%%%%%%%%%%%%%%%%%%%%%%%%%%%%%%%%%%%%%%%%%%%%%%%%%%%%%%
%%%%%%%%%%%%%%%%%%%%%%%%%%%%%%%%%%%%%%%%%%%%%%%%%%%%%%%%%%%%%%%%%%%%%%%%%%%%%%%%%
\clearpage

\clearpage
%%%%%%%%%%%%%%%%%%%%%%%%%%%%%%%%%%%%%%%%%%%%%%%%%%%%%%%%%%%%%%%%%%%%%%%%%%%%%%%%%
%%%%%%%%%%%%%%%%%%%%%%%%%%%%%%%%%%%%%%%%%%%%%%%%%%%%%%%%%%%%%%%%%%%%%%%%%%%%%%%%%

%%%%%%%%%%%%%%%%%%%%%%%
%%%%%%%%%%%%%%%%%%%%%%%%%%%%%%%%%

%%%%%%%%%%%%%%%%%%%%%%%%%%%%%%%%%
%%%%%%%%%%%%%%%%%%%%%%%%%%%%%%%%%
%%%%%%%%%%%%%%%%%%%%%%%%%%%%%%%%%

\newpage 

\newpage

\section{APPENDIX}
%%%%%%%%%%%%%%%%%%%%%%%%%%%%%%%%%

\subsection{On the 5hmC measure $\Delta \beta(\alpha)$}
%%%%%%%%%%%%%%%%%%%%%%%%%%%%%%%%%
\subsubsection{Basic properties of $\Delta \beta(\alpha)$}\label{sec2}

%%%%%%%%%%%%%%%%%%%%%%%%%%%%%
%%%%%%%%%%%%%%%%%%%%%%%%%%%%%
In this section we summarize basic properties of the measure $\Delta \beta(\alpha)$ defined as 
\begin{eqnarray}\label{b11}
\Delta \beta(\alpha) = \beta_{BS} - \beta_{oxBS}=\frac{M_{BS}}{M_{BS}+U_{BS}+\alpha} - \frac{M_{oxBS}}{M_{oxBS}+U_{oxBS}+\alpha},
\end{eqnarray}
with a correction term $\alpha >0$.  
\vspace{0.3cm}
\\
First,  $\Delta \beta(\alpha)$ in (\ref{b11})   is well-defined for any  $M_i, U_ i \geq 0,\,\, i \in \{BS, oxBS\}$ and $\alpha > 0$  and ranges between -1 and 1. 
\vspace{0.3cm}
\\
\noindent Since the condition $\{\Delta \beta(\alpha) >0\}$ plays a key role in the screening step, we need to determine when this condition will hold. Standard calculations show that  $\{\Delta \beta(\alpha) >0\}$ is equivalent to the condition
\begin{eqnarray}\label{nm}
\frac{M_{BS}}{M_{oxBS}} > \frac{U_{BS}+\alpha}{U_{oxBS}+\alpha}.
\end{eqnarray}
Thus all CpGs  satisfying (\ref{nm}) will also satisfy $\{\Delta \beta(\alpha) >0\}.$
\vspace{0.4cm}
\noindent Further, to capture the behaviour of $\Delta \beta(\alpha)$ for increasing values of $\alpha$, we take the limit in (\ref{b11}) and obtain
\begin{eqnarray}\label{iii}
\lim_{\alpha \uparrow \infty} \Delta \beta (\alpha) = 0,
\end{eqnarray}
for a given sample and CpG.

%%%%%%%%%%%%%%%%%%%%%%%%%%%%%%%
\subsubsection{$\Delta \beta(\alpha)$ as a function of $\alpha$}\label{a1}
{\bf Maxima / minima of the function $\Delta \beta(\alpha)$}
\vspace{0.15cm}
\\
\noindent To address possible impact of $\alpha$ on the values of $\Delta \beta(\alpha)$ analytically, we regard  $\Delta \beta$ as a function of $\alpha$ and compute the corresponding derivative
\begin{eqnarray}
\frac{d \Delta \beta(\alpha)}{d \alpha} = \frac{M_{oxBS}}{(M_{oxBS}+U_{oxBS}+\alpha)^2} - \frac{M_{BS}}{(M_{BS}+U_{BS}+\alpha)^2}.
\\ \nonumber
\end{eqnarray}

\noindent This derivative becomes zero in
\begin{eqnarray}\label{a0}
\alpha_0 = \frac{M_{oxBS}\,\sqrt{M_{BS}}+U_{oxBS}\,\sqrt{M_{BS}}-M_{BS}\,\sqrt{M_{oxBS}}-U_{BS}\,\sqrt{M_{oxBS}}}{\sqrt{M_{oxBS}}-\sqrt{M_{BS}}}.
\end{eqnarray}
The quantity in (\ref{a0}) is well-defined for all CpGs with $M_{BS}\neq M_{oxBS}$ and satisfies $\alpha_0 > 0$ either if\footnote{For $\alpha_0 < 0$ we just consider $\hat{\alpha_0} =0$ and compute $$\Delta \beta(0) = \frac{M_{BS} U_{oxBS} - M_{oxBS} U_{BS}}{(M_{BS}+U_{BS} )(M_{oxBS}+U_{oxBS})}.$$  This quantity is positive for all CpGs satisfying $$\frac{M_{BS}}{M_{oxBS}} > \frac{U_{BS}}{U_{oxBS}} \,\,\,\text{or, equivalently,} \,\,\,\frac{M_{BS}}{M_{oxBS}} > \frac{M_{BS}+U_{BS}}{M_{oxBS}+U_{oxBS}}.$$} 
\begin{eqnarray}\label{a01}
M_{BS} > M_{oxBS} \,\,\,\,\,\text{and}\,\,\,\,\,\,\, \frac{M_{BS}+U_{BS}} {M_{oxBS}+U_{oxBS}} > \sqrt{\frac{M_{BS}}{M_{oxBS}}}
\end{eqnarray}
or if 
\begin{eqnarray}\label{a02}
M_{BS} < M_{oxBS}\,\,\,\,\,\text{and}\,\,\,\,\,\,\,  \frac{M_{BS}+U_{BS}} {M_{oxBS}+U_{oxBS}} < \sqrt{\frac{M_{BS}}{M_{oxBS}}}. 
\end{eqnarray}
\vspace{-0.5cm}
\\
\\
Further,   $\alpha_0$ will be a minimum of $\Delta \beta(\alpha)$ if $M_{oxBS}+U_{oxBS} > M_{BS}+U_{BS}$, and its maximum  if $M_{oxBS}+U_{oxBS} < M_{BS}+U_{BS}$.
\vspace{0.4cm}
\\
The function value $\Delta \beta(\alpha_0)$ for $\alpha_0$ as in (\ref{a0}) is given by 
\begin{eqnarray}
\Delta \beta(\alpha_0) = \frac{(\sqrt{M_{BS}}-\sqrt{M_{oxBS}})^2}{M_{BS}+U_{BS}-M_{oxBS}-U_{oxBS}}
\end{eqnarray}
and will be positive for all CpGs with $M_{BS}+U_{BS} > M_{oxBS}+U_{oxBS}$  and negative for all CpGs with $M_{BS}+U_{BS}<M_{oxBS}+U_{oxBS}$. 
\vspace{0.3cm}
\\
Given that the conditions  (\ref{a01}) imply $M_{BS}+U_{BS} > M_{oxBS}+U_{oxBS}$ and the conditions (\ref{a02}) imply $M_{BS}+U_{BS}<M_{oxBS}+U_{oxBS}$, we summarize that $\alpha_0$ satisfying (\ref{a01})  represents the maximum of the function $\Delta \beta(\alpha)$, with $\Delta \beta(\alpha_0)>0$;  on the other hand, $\alpha_0$ satisfying (\ref{a02})  is the minimum of the function $\Delta \beta(\alpha)$, with $\Delta \beta(\alpha_0)<0$. 
\vspace{0.35cm}
\\
{\bf Monotonicity of $\Delta \beta(\alpha)$}
\vspace{0.15cm}
\\
One can easily verify that for $M_{oxBS}>M_{BS}$ the function $\Delta \beta(\alpha)$ will be increasing  for all $\alpha$ in the interval  $ (\alpha_0, \infty)$;
 for $M_{oxBS} < M_{BS}$  this function will be increasing for all $\alpha$ in the interval $(-\infty, \alpha_0).$

%%%%%%%%%%%%%%%%%%%%%%%%%%%%%%%% 

 \subsubsection{Sign change of $\Delta \beta(\alpha)$ }\label{a2}
 \noindent Standard calculations show that $\Delta \beta(\alpha)$ may change its sign either for the CpGs with 
\begin{eqnarray}\label{jk_b}
\frac{M_{BS}}{M_{oxBS}} > \frac{M_{BS}+U_{BS}}{M_{oxBS}+U_{oxBS}}\,\,\,\,\, \text{and} \,\,\,\,\, M_{oxBS} > M_{BS}
\end{eqnarray}
 or those with 
 \begin{eqnarray}\label{jk2_b}
 \frac{M_{BS}}{M_{oxBS}} < \frac{M_{BS}+U_{BS}}{M_{oxBS}+U_{oxBS}}\,\,\,\,\, \text{and} \,\,\,\,\,M_{oxBS} < M_{BS}.
 \end{eqnarray}
\\
We can also compute the actual value $\alpha^* > 0$, with  $\Delta \beta(\alpha^*) =0$ and $\Delta \beta(\alpha)$ changing its sign in $\alpha^*$\footnote{Simple calculations show that $\alpha_0 > \alpha^*$ for all CpGs satisfying both inequalities $M_{oxBS}+U_{oxBS} > M_{BS}+U_{BS}$ and $M_{oxBS} >M_{BS}$  simultaneously or those with $M_{oxBS}+U_{oxBS} < M_{BS}+U_{BS}$ and $M_{oxBS} < M_{BS}$. 
}. Indeed, from (\ref{b11}) it follows that  $\Delta \beta (\alpha^*) = 0$ for 
\begin{eqnarray}\label{jue2}
\alpha^* = \frac{M_{oxBS} U_{BS} - M_{BS} U_{oxBS}}{M_{BS} -M_{oxBS}}.
\end{eqnarray}
Note that both (\ref{jk_b}) and (\ref{jk2_b}) guarantee that $\alpha^* > 0.$ 
\vspace{0.3cm}
\\
Further, the condition (\ref{a02}) leads to the condition (\ref{jk_b}) and thus for all CpGs satisfying (\ref{a02}) the measure $\Delta \beta(\alpha)$ will indeed  change its sign in $\alpha^*$; the same result holds for the conditions (\ref{a01}) and (\ref{jk2_b}).

%%%%%%%%%%%%%%%%%%%%%%%%%%%%%%%%%%%%%%%%%%%%%%%%%%%%%%%%%%%%%%%%%%%%%%%%
%%%%%%%%%%%%%%%%%%%%%%%%%%%%%%%%%%%%%%%%%%%%%%%%%%%%%%%%%%%%%%%%%%%%%%%%

\subsection{On the 5hmC measures $\Delta m (\alpha)$ and $\Delta m^{\infty}$}
%%%%%%%%%%%%%%%%%%%%%%%%%%%%%%%%%

\subsubsection{Basic properties} \label{wd}

When defined by 
\begin{eqnarray}
\Delta m = m_{BS} - m_{oxBS} =\log_2 \frac{\beta_{BS}}{1-\beta_{BS}} -\log_2\frac{\beta_{oxBS}}{1-\beta_{oxBS}},
\end{eqnarray}
the measure $\Delta m$ is well-defined  for any $M_i, U_ i \geq 0,\,\, i \in \{BS, oxBS\},$ if $\alpha > 0$. For $\alpha = 0 $ $\Delta m$ is also well-defined, but only if $M_i $ and $U_i$  do not vanish simultaneously. This result conforms to the fact that the measure $\Delta \beta(\alpha)$ is  well-defined for any  $M_i, U_ i \geq 0,\,\, i \in \{BS, oxBS\}$ as $\alpha > 0$; for $\alpha = 0$,  $M_{BS}, M_{oxBS}, U_{BS}, U_{oxBS}$ must not vanish at the same time for $\Delta \beta(\alpha)$ to be well-defined. Thus the 5hmC measures $\Delta \beta(\alpha)$ and $\Delta m$ are well-defined on the same sets of CpGs.
\vspace{0.3cm}
\\
For the measure $\Delta m (\alpha)$ calculated as 
\begin{eqnarray}\label{ui3}
\Delta m(\alpha)=\log_2\frac{M_{BS}}{M_{oxBS}}+\log_2\frac{U_{oxBS} +\alpha}{U_{BS}+\alpha}. 
\end{eqnarray}
to be well-defined, we have to differentiate between the following two situations:
\begin{enumerate}
\item $\alpha >0$. In these cases  the conditions
\begin{eqnarray}\label{ui22}
M_{BS}\neq 0 \,\,\, \text{and} \,\,\, M_{oxBS}\neq 0
\end{eqnarray} 
must hold simultaneously, since otherwise the term $\log_2 \frac{M_{BS}}{M_{oxBS}}$ in (\ref{ui3}) will become infinite. 
\item $\alpha=0$. In these cases the conditions
\begin{eqnarray}\label{ui33}
M_{BS}\neq 0,  M_{oxBS}\neq 0, U_{BS}\neq 0, U_{oxBS}\neq 0
\end{eqnarray}
must hold simultaneously, since otherwise both terms  $\log_2 \frac{M_{BS}}{M_{oxBS}}$ and $\log_2 \frac{U_{BS}+\alpha}{U_{oxBS}+\alpha}$ in (\ref{ui3}) will become infinite. 
\end{enumerate}
\noindent Given that only a few CpGs in our real data sets do not satisfy these conditions, we used the definition (\ref{ui3}) of $\Delta m$ in all our discussions. 
\\
\\
\vspace{0.15cm}
\noindent  \noindent  Further, the condition $\Delta m(\alpha) > 0$ is equivalent to the inequality
\begin{eqnarray}\label{ineq}
\frac{M_{BS}}{M_{oxBS}} > \frac{U_{BS}+\alpha}{U_{oxBS}+\alpha}.
\end{eqnarray}
This inequality is exactly the same as the inequality (\ref{nm}) that has been shown to imply $\Delta \beta(\alpha) >0.$ Thus, for any fixed values of $M_{BS}, \,M_{oxBS},\, U_{BS}$ and  $U_{oxBS}$  the measures  $\Delta m(\alpha)$ and $\Delta \beta(\alpha)$ must have the same sign. Indeed, 
$\Delta m > 0$ implies $$\frac{\beta_{BS}}{\beta_{oxBS}} \cdot \frac{1-\beta_{oxBS}}{1-\beta_{BS}} > 1$$ which is equivalent to $\beta_{BS} > \beta_{oxBS}$ or, alternatively,  $\Delta \beta >0$; on the other hand, $\Delta \beta > 0$ leads to the inequality (\ref{ineq}) and thus $\Delta m > 0$ follows. 
\vspace{0.3cm}
\\
\noindent Similarly to the case of $\Delta \beta(\alpha)$, the convergence of $\Delta m(\alpha)$ for increasing $\alpha$ can easily be derived from (\ref{ui3}), which gives
\begin{eqnarray}\label{conv}
\lim_{\alpha \uparrow \infty} \Delta m ( \alpha) = \log_2 \frac{M_{BS}}{M_{oxBS}}.
\end{eqnarray}
Since  $\Delta m(\alpha)$  is monotone in $\alpha$ (see below), we can consider $ \log_2 \frac{M_{BS}}{M_{oxBS}}$  as an upper or lower bound for $\Delta m(\alpha)$, depending on whether $U_{BS} > U_{oxBS}$ or $U_{BS} < U_{oxBS}.$

%%%%%%%%%%%%%%%%%%%%%%%%%%%%%%%
\subsubsection{$\Delta m(\alpha)$ versus  $\Delta \beta(\alpha)$: no functional dependence}\label{fgh77}
\begin{figure}[htp]
    \hspace{2.1cm} \includegraphics[width=0.7\linewidth, height=0.21\textheight]{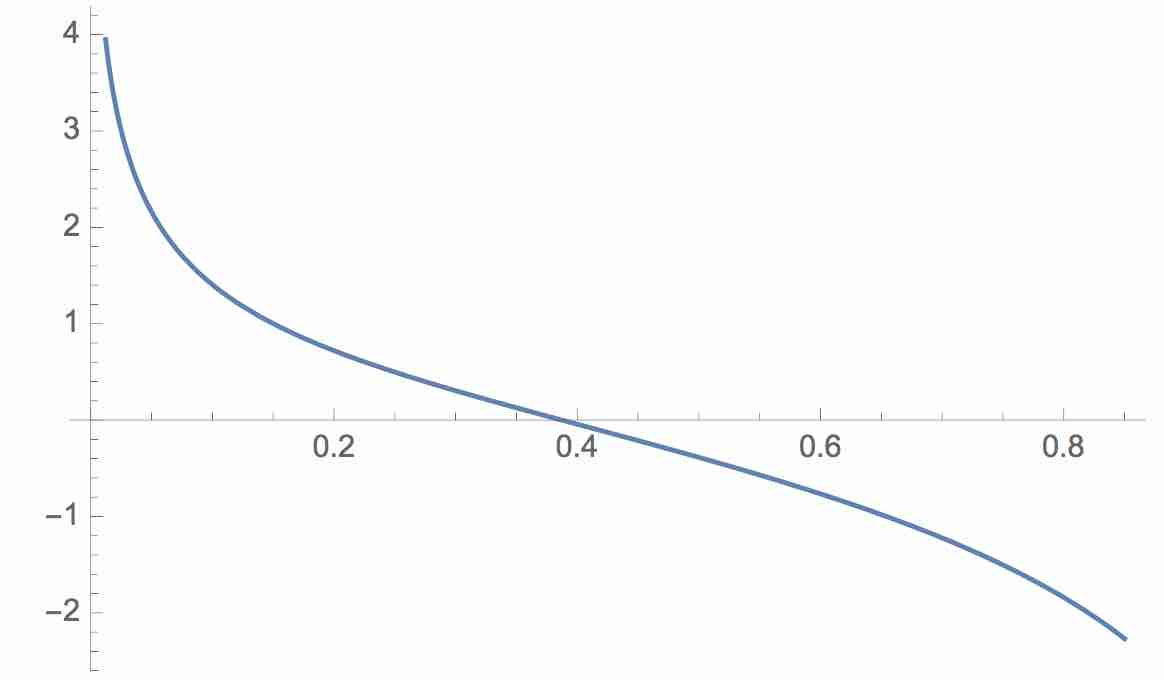}
        \caption{$\Delta m(\alpha)$ as a function of $\beta_{oxBS}$. The x-axis corresponds to the $\beta_{oxBS}(\alpha)$ values and the y-axis corresponds to the $\Delta m(\alpha)$ values. Note that in this example $\Delta \beta(\alpha)$ is constant for all values of $\beta_{oxBS}(\alpha)$.}
        \label{wsc}
 \end{figure}

In this section we show that there is no transformation between the measures $\Delta m(\alpha)$ and $\Delta \beta(\alpha)$ that would render $\Delta m(\alpha)$ as a function of $\Delta\beta(\alpha)$. To see this,  we consider an example with $\beta_{oxBS}$ ranging between $0$ and $0.85$ and $\beta_{BS}=\beta_{oxBS}+0.15.$ In such setting, $\Delta \beta(\alpha)$ will be constant, but $\Delta m(\alpha)$ will change its values for changing $\beta_{oxBS}(\alpha)$; thus $\Delta m(\alpha) $ is not a function of $\Delta \beta(\alpha)$. For an illustration of this result see Figure \ref{wsc}.
\vspace{-0.2cm}

\clearpage

%%%%%%%%%%%%%%%%%%%%%%%%%%%%%%%
\subsubsection{$\Delta m(\alpha)$ as a function of $\alpha$}\label{fgh}

We consider $\Delta m (\alpha)$ as a function of $\alpha$ and compute the derivative
\begin{eqnarray}\label{deriv}
\frac{d\Delta m(\alpha)}{d \alpha} = \frac{U_{BS}-U_{oxBS}}{ (U_{BS} + \alpha) (U_{oxBS} + \alpha)\,\ln 2}
\end{eqnarray} 
This derivative is obviously positive for all CpGs with $U_{BS} \geq  U_{oxBS}$; thus for such CpGs  $\Delta m(\alpha)$ will increase with increasing $\alpha$. On the other hand, for all CpGs with $U_{BS} < U_{oxBS}$  $\Delta m(\alpha)$  is decreasing in $\alpha$. 
\vspace{0.2cm}
\\
Since on the one hand the derivative (\ref{deriv}) does not have zeros and on the other hand $\alpha \geq 0$ must hold, we set the minimum of  $\Delta m(\alpha)$ to be attained at $\alpha = 0$ for all CpGs with $U_{BS} \geq U_{oxBS}$ ; the increasing values of $\alpha$ in this case will lead to increasing values of  $\Delta m(\alpha)$. 
\vspace{0.3cm}
\\For the CpGs with $U_{BS} < U_{oxBS}$, the maximal  value of $\Delta m(\alpha)$ will be attained at $\alpha = 0$; the increasing values of $\alpha$ in this case will correspond to  decreasing values of  $\Delta m(\alpha)$. This last fact also implies that all CpGs with $U_{BS} < U_{oxBS}$ and  $\Delta m(0) < 0$ should be excluded from the data set immediately as irrelevant for the further analysis, even before the screening step is completed.
\vspace{0.2cm}

%%%%%%%%%%%%%%%%%%%%%%%%%%%%%%%
%%%%%%%%%%%%%%%%%%%%%%%%%%%%%%%

\subsubsection{Sign change of $\Delta m(\alpha)$ }\label{a22}
Standard calculations show that $\Delta m (\alpha)$ changes its sign either for the CpGs with  
\begin{eqnarray}\label{e15}
M_{BS} > M_{oxBS}\,\,\,\,\,\text{and} \,\,\,\,\,\frac{M_{BS}+U_{BS}}{M_{oxBS}+U_{oxBS}} > \frac{M_{BS}}{M_{oxBS}}
\end{eqnarray}
or those with
\begin{eqnarray}\label{e16}
M_{BS} < M_{oxBS}\,\,\,\,\,\text{and}\,\,\,\,\,\frac{M_{BS}+U_{BS}}{M_{oxBS}+U_{oxBS}} < \frac{M_{BS}}{M_{oxBS}}.
\end{eqnarray}  
Note that these are exactly the conditions (\ref{jk_b}) and (\ref{jk2_b}) that imply the sign changes for the measure $\Delta \beta(\alpha)$. 
\vspace{0.2cm}
\\
Here we also compute the value $\alpha^* > 0$ for which  $\Delta m(\alpha^*) =0$ and $\Delta m(\alpha)$ changes its sign at $\alpha^*.$ Indeed, from (\ref{ui3}) it follows that  $\Delta m (\alpha^*) = 0$ for 
\begin{eqnarray}\label{ju}
\alpha^* = \frac{M_{oxBS} U_{BS} - M_{BS} U_{oxBS}}{M_{BS} -M_{oxBS}},
\end{eqnarray}
which is exactly the same $\alpha^*$  as obtained in (\ref{jue2})  for $\Delta \beta(\alpha).$

%%%%%%%%%%%%%%%%%%%%%%%%%%%%%%

\subsubsection{On the relation between the subsets $\{\Delta m(\alpha) >0\}$ and $\{\Delta m^{\infty}>0\}$}\label{infty}
The aim of this section is to show analytically that, for a given sample and increasing $\alpha>0$, the subset of CpGs satisfying $\Delta m^{\infty}>0$ will approximately become a "limiting" subset for a sequence of subsets  of CpGs satisfying $\Delta m(\alpha)>0$. The discussion above should hold for a given sample. 
\vspace{0.25cm}
\\
First, let us assume that $\Delta m^{\infty}>0$ for a given sample and CpG. Then we want to show that for $\alpha>0$ large enough, it will hold $\Delta m(\alpha)>0$ for this sample and CpG, too.  The condition $\Delta m^{\infty}>0$ evidently leads to $M_{BS}>M_{oxBS}.$ On the other hand, while considering the second term in the expression $$\Delta m(\alpha)=\log_2\frac{M_{BS}}{M_{oxBS}}+\log_2 \frac{U_{oxBS}+\alpha}{U_{BS}+\alpha},$$
we state that 
\begin{eqnarray}\label{uio}
\forall \epsilon >0 \,\,\text{exists a }\,\, \alpha_0>0\,\,\text{such that}\,\, \Big| \log_2 \frac{U_{oxBS}+\alpha}{U_{BS}+\alpha}\Big| < \epsilon. 
\end{eqnarray}
Thus for all $\alpha>\alpha_0$ we will get $\Delta m(\alpha)>0.$
\vspace{0.25cm}
\\
Next, let us assume that $\Delta m(\alpha)>0$ for all $\alpha>0$ large enough. Then, with \eqref{uio}, it must hold $$\log_2\frac{M_{BS}}{M_{oxBS}}\geq 0.$$ This latter expression implies either $M_{BS}>M_{oxBS}$, and, as a result, $\Delta m^{\infty}>0$, or $M_{BS}=M_{oxBS}.$\footnote{As a reminder: for $M_{BS}=M_{oxBS}$ the condition $\Delta m(\alpha)>0$ will hold only if $U_{BS}<U_{oxBS}$. But, due to our discussion above, we agreed to ignore the set $\{M_{BS}=M_{oxBS}, U_{BS}<U_{oxBS}\}$.}
\vspace{0.35cm}
\\
Similar considerations show that in cases when $\alpha>0$ increases, the subset of CpG sites satisfying $\{\Delta m(\alpha)>0\}$ over all given samples (approximately) approaches the subset of CpG sites which satisfy $\Delta m^{\infty}>0$ over all given samples; for an illustration see Figure \ref{sub3335}.

%%%%%%%%%%%%%%%%%%%%%%%%%%%%%%%%%
\subsection{On the 5hmC measure $\Delta h$}
%%%%%%%%%%%%%%%%%%%%%%%%%%%%%%%%%

\subsubsection{On the relation between the subsets $\{\Delta h >0\}$, $\{\Delta m(\alpha) >0\}$ and $\{\Delta m^{\infty}>0\}$} \label{wd100}
In this section we analyze how the positivity of the 5hmC  $\Delta h$ implies the positivity of two other 5hmC measures.  First of all, as follows from the definition of $\Delta h$, the inequality $\Delta h >0$ holds for all CpGs with  $$M_{BS}+U_{BS} > M_{oxBS}+U_{oxBS}.$$ Further, if we have the inequalities $M_{BS}+U_{BS} > M_{oxBS}+U_{oxBS}$ and $M_{BS}> M_{oxBS}$ holding simultaneously, both measures $\Delta h$ and $\Delta m^{\infty}$ will be positive. 
\vspace{0.3cm}
\\
For CpGs satisfying
 $$M_{BS}+U_{BS} > M_{oxBS}+U_{oxBS}, \,M_{BS} > M_{oxBS},\,\,\text{and}\,\, U_{BS} < U_{oxBS}$$
 the 5hmC measure $\Delta m(\alpha)$ will be positive for any $\alpha>0.$ If $U_{BS}>U_{oxBS}$, then the sign of $\Delta m(\alpha)$ will depend on the choice of $\alpha$. In particular, we will have $\Delta m(\alpha) >0$ for $\alpha>\alpha^*$, with $\alpha^*$ as in (\ref{ju}); otherwise $\Delta m(\alpha)$ will be negative.
\vspace{0.3cm}
\\
For CpGs with $$M_{BS}+U_{BS} > M_{oxBS}+U_{oxBS}\quad\text{and}\quad M_{BS}< M_{oxBS} \quad\text{and}\quad U_{BS} > U_{oxBS},$$
 we will get $\Delta h>0$, but the measures $\Delta m^{\infty}$ and $\Delta m(\alpha)$ will have negative outcomes.

%%%%%%%%%%%%%%%%%%%%%%%%%%%%%%%%
%%%%%%%%%%%%%%%%%%%%%%%%%%%%%%%%


\newpage
\begin{thebibliography}{99}


\bibitem{Booth1} Booth M. J., Ost T.W.B., Beraldi D., Bell N. M., Branco M. R., Reik W., Balasubramanian S. {\sl Oxidative bisulfite sequencing of 5-methylcytosine and 5-hydroxymethylcytosine} 
 Net Protoc. 2013 October; 8(10): 1841-1851.

\bibitem{li} Li D., Xie Z., Pape M. L., Dye T. {\sl An evaluation of statistical methods for DNA methylation microarray data analysis} 
BMC Bioinformatics (2015) 16.


\bibitem{stewart} Stewart S. K., Morris T. J., Guilhamon P., Bulstrode H., Bachman M., Balasubramanian S., Beck S.  {\sl oxBS-450K: a method for analysing hydroxymethylation using 450K BeadChips} 
Methods 72 (2015) 9-15.

\bibitem{field} Field S. F., Beraldi D., Bachman M., Stewart S. K., Beck S., Balasubramanian S.   {\sl Accurate measurement of 5-methylcytosine and 5-hydroxymethylcytosine in human cerebellum DNA by oxidative bisulfite on an array (oxBS-array} PLoS ONE 10(2) (2015)

\bibitem{godderis} Godderis L., Schouteden C., Tabish A., Poels K., Hoet P., Baccareli A. A., Van Landuyt K.   {\sl Global methylation and hydroxymethylation in DNA from blood and saliva in healthy Volunteers} 
Hindawi Publishing Corporation, BioMed Research International, Volume 2015, Article ID 845041.

\bibitem{yu} Yu M., Hon G. C., Szulwach K.E., Song C.-X., Zhang L., Kim A., Li X., Dai Q., Park B., Min J.-H., Jin P., Ren B., He C. {\sl Base-Resolution Analysis of 5-Hydroxymethylcytosine in the Mammalian Genome } 
Cell. 2012 June 8; 149(6): 1368-1380.

\bibitem{nazor} Nazor K.L., Boland M. J., Bibikova M., Klotzle B., Yu M., Glenn-Pratola V. L., Schell J. P., Coleman R. L., Cabral-da-Silva M. C., Schmidt U., Peterson S. E., He C., Loring J. F., Fan J.-B. {\sl Application of a low cost array-based technique - TAB -Array-for quantifying and mapping  both 5mC and 5hmC at single base resolution in human pluripotent stem cells } 
Genomics 104 (2014) 358-367.

\bibitem{nestor} Nestor C. E., Ottaviano R., Reddington J., Sproul D., Reinhardt D., Dunican D., Katz E., Dixon J.M., Harrison D. J., Meehan R.R.{\sl Tissue type is a major modifier of the 5-hydroxymethylcytosine content of human genes. } 
Genome Research 22: 467-477.


\bibitem{bib}  Bibikova M., Barnes B., Tsan C., Ho V., Klotzle B., Le J. M., Delano D., Zhang L., Schroth G.P., Gunderson K. L., Fan J.-B., Shen R. {\sl High density DNA methylation array with single CpG site resolution} 
Genomics 98 (2011) 288-295.

\bibitem{pan} Pan D., Zhang X., Huang C. - C., Jafari N., Kibbe W. A., Hou L., Lin S. M.  {\sl Comparison of Beta - value and M - value methods for quantifying methylation levels by microarray analysis} 
BMC Bioinformatics 2010, 11:587.

\bibitem{housemann} Houseman E. A., Johnson K. C., Christensen B. C.  {\sl OxyBS: estimation of 5-methylcytosine and 5-hydroxymethylcytosine from tandem-treated oxidative bisulfite and bisulfite DNA}  BMC Bioinformatics 2016, 1-3.

\bibitem{tellez} Tellez-Plaza M., Tang W., Shang Y., Umans J. G., Francesconi K. A., Goessler W., Ledesma M., Leon M., Laclaustra M., Pollak J., Guallar E., Cole S. A., Fallin M. D., Navas-Acien A. {\sl Association of global DNA methylation and global DNA hydroxymethylation  with metals and other exposures in human blood DNA samples.}  Environ Health Perspect 122: 946-954.

\bibitem{bachman} Bachman A., Uribe-Lewis S., Yang X., Williams M., Murrell A., Balasubramanian S. {\sl 5-Hydroxymethylcytosine is a predominantly stable DNA modification.}  NATURE Chemistry, Vol. 6, December 2014

\bibitem{geest} P. A. G. van der Geest {\sl An algorithm to generate samples of multi-variate distributions with correlated marginals.}  Computational Statistics and Data Analysis, Volume 27 Issue 3, May 1998

\bibitem{laird} A. Laird, J. P. Thomson, D.J. Harrison, R.R. Meehan {\sl 5-hydroxymethylcytosine profiling as an indicator of cellular state.}  Epigenomics (2013) 5(6), 655-669.

\bibitem{diserud} O. H. Diserud, F.  Odegaard {\sl A multiple-site similarity measure.}  Biol. Lett. (2007) 3, 20-22

\end{thebibliography}
\end{document}